\documentclass[pra,twocolumn,preprintnumbers,amsmath,amssymb,superscriptaddress]{revtex4-1}

\usepackage{graphicx}
\usepackage{dcolumn}
\usepackage{bm}

\usepackage[T1]{fontenc}
\usepackage[latin1]{inputenc}
\usepackage[english]{babel}
\usepackage{amsmath}
\usepackage{amssymb}
\usepackage{amstext}
\usepackage{bbold}
\usepackage{braket}
\usepackage{relsize}
\usepackage{subfigure}
\usepackage{hhline}
\usepackage{booktabs}

\newcommand{\eq}[1]{\begin{equation} #1 \end{equation}}
\newcommand{\eqa}[1]{\begin{eqnarray} #1 \end{eqnarray}}

\newcommand{\be}{\begin{equation}}
\newcommand{\ee}{\end{equation}}
\newcommand{\ba}{\begin{eqnarray}}
\newcommand{\ea}{\end{eqnarray}}

\newcommand{\ue}{\mathrm{e}}

\begin{document}

\preprint{}

\title{
Can One Trust Quantum Simulators?}
\author{Philipp Hauke}
    \email{philipp.hauke@icfo.es}
    \affiliation{ICFO -- Institut de Ci\`{e}ncies Fot\`{o}niques, Parc Mediterrani de la Tecnologia, 08860 Castelldefels, Spain}
\author{Fernando M.\ Cucchietti}
    \affiliation{ICFO -- Institut de Ci\`{e}ncies Fot\`{o}niques, Parc Mediterrani de la Tecnologia, 08860 Castelldefels, Spain}
    \affiliation{Barcelona Supercomputing Center (BSC-CNS), Edificio NEXUS I, Campus Nord UPC, Gran Capit\'{a}n 2-4, 08034 Barcelona, Spain}
\author{Luca Tagliacozzo}
    \affiliation{ICFO -- Institut de Ci\`{e}ncies Fot\`{o}niques, Parc Mediterrani de la Tecnologia, 08860 Castelldefels, Spain}
\author{Ivan Deutsch}
    \affiliation{Center for Quantum Information and Control (CQuIC)}
    \affiliation{Department of Physics and Astronomy, University of New Mexico, Albuquerque NM 87131}
\author{Maciej Lewenstein}
    \affiliation{ICFO -- Institut de Ci\`{e}ncies Fot\`{o}niques, Parc Mediterrani de la Tecnologia, 08860 Castelldefels, Spain}
    \affiliation{ICREA -- Instituci{\'o} Catalana de Recerca i Estudis Avan\c{c}ats, Lluis Companys 23, E-08010 Barcelona, Spain}
\date{\today}

\begin{abstract}
Various fundamental phenomena of strongly-correlated quantum
systems such as high-$T_c$ superconductivity, the fractional
quantum-Hall effect, and quark confinement are still awaiting a
universally accepted explanation.  The main obstacle is the
computational complexity of solving even the most simplified
theoretical models that are designed to capture the relevant
quantum correlations of the many-body system of interest.   In his
seminal 1982 paper \cite{Feynman1982}, Richard Feynman suggested
that such models might be solved by ``simulation'' with a new type
of computer whose constituent parts are effectively governed by a
desired quantum many-body dynamics. Measurements on this
engineered machine, now known as a ``quantum simulator,'' would
reveal some unknown or difficult to compute properties of a model
of interest. 
We argue that a useful quantum simulator must satisfy four
conditions: \textit{relevance}, \textit{controllability},
\textit{reliability}, and \textit{efficiency}. We review the
current state of the art of {\em digital} and {\em analog}
quantum simulators. 
Whereas so far the majority of the focus, both
theoretically and experimentally, has been on controllability of relevant models, we emphasize 
here 
the need for a careful analysis of reliability and efficiency in the presence of imperfections.  
We discuss how disorder and noise can impact these conditions, and illustrate our concerns with 
novel numerical simulations of a paradigmatic example: a disordered
quantum spin chain governed by the Ising model in 
a transverse magnetic field. We find that disorder can decrease the reliability of an
analog quantum simulator of  this model, although large errors
in local observables are introduced only for strong levels of
disorder. We conclude that the answer to the question ``Can we
trust quantum simulators?'' is... to some extent.
\end{abstract}


\maketitle

\section{Introduction}

In his 1982 foundational article \cite{Feynman1982}, Richard Feynman
suggested that the complexities of quantum many-body physics might
be computed by ``simulation.''  By designing a well-controlled
system from the bottom up, one could create a computer whose
constituent parts are governed by quantum dynamics generated by a
desired Hamiltonian. Measuring the properties of this
nano-engineered system thus reveals some unknown or difficult to
compute properties of a quantum many-body model, such as the
nature of quantum phase diagrams.  Feynman's machine is now known
as a ``quantum simulator'' (QS).

Fueled by the prospect of solving a broad range of long-standing
problems in strongly-correlated systems, the tools to design,
build, and implement QSs
\cite{Feynman1982,Buluta2009,Lewenstein2012} have rapidly developed and
are now reaching very sophisticated levels~\cite{RunnersUp2010}.
Researchers are making breakthrough advances in quantum control of
a variety of systems, including ultracold atoms and molecules
(\cite{Lin2009b,Bakr2010,Trotzky2010,Jordens2010,Baumann2010,Liao2010,Simon2011,Struck2011a,VanHoucke2012,Trotzky2012,Cheneau2012,Aidelsburger2011,Jimenez2012},
 for reviews see also \cite{Bloch2008,Bloch2012,Mueller2012}), ions
(\cite{Friedenauer2008,Kim2010,Islam2011,Lanyon2011,Barreiro2011,Gerritsma2010,Gerritsma2011},
for recent reviews see \cite{Johanning2009,Mueller2012,Schneider2012,Blatt2012}),
photons
(\cite{Lanyon2007,Lanyon2008,Lanyon2010,Belgiorno2010,Kitagawa2011,Ma2011}, for a recent review see \cite{AspuruGuzik2012}), 
circuit quantum electrodynamics (CQED) and polaritons
(\cite{Greentree2006,Hartmann2006,Angelakis2007,Wang2009a,Koch2010},
for a recent review see \cite{Angelakis2012}), artificial lattices
in solid state \cite{Singha2011}, nuclear magnetic resonance
(NMR) systems \cite{Somaroo1999,Tseng1999,Du2010,Alvarez2010,Li2011,Lu2011}, and superconducting qubits (for reviews see \cite{Makhlin2001,Devoret2004,You2005,Girvin2009}, for the current state of art \cite{Pritchett2010,Reed2012} and references therein). For a general overview see also \cite{Zoller2005}.
At the current pace, it
is expected that we will soon reach the ability to finely control
many-body systems whose description is outside the reach of a
classical computer. For example, modeling interesting physics
associated with a quantum system involving 50 spin-1/2 particles
-- whose general description requires $2^{50} \approx 10^{15}$
complex numbers -- is out of the reach of current classical
supercomputers, but perhaps within the grasp of a QS.

In a field brimming with excitement, it is important to critically
examine such high expectations.  Real-world implementations of a
quantum simulation will always face experimental imperfections,
such as noise due to finite precision instruments and interactions
with the environment.  
Feynman's QS is often considered as a fundamentally {\em analog} device, in the sense that all operations are carried out continuously.  
However, errors in an analog device (also continuous, like temperature in the initial
state, or the signal-to-noise ratio of measurement) can propagate
and multiply uncontrollably~\cite{Kendon2010}.  Indeed, Landauer, a
father of the studies of the physics of information, questioned
whether quantum coherence was truly a powerful resource for
computation because it required a continuum of possible superposition
states that were ``analog'' in nature~\cite{Landauer1996}.

This contrasts with the operation of a universal {\em digital}
quantum computer as envisioned by David Deutsch, in which all
operations are digitized into a finite set of logic gates and
measurements~\cite{Deutsch1985} \footnote{Note that we use the notion of digital quantum simulator in this sense, i.e., the digitization of operations, not in the sense of digits of precision. In particular, the outcome of an analog quantum simulator will also yield only a finite digital precision, but its operations are performed continuously.}.  The invention of quantum-error-correcting codes showed that a quantum computer is in some sense
{\em both} analog and digital.  Through a discrete set of unitary
transformations, we can get arbitrarily close to any
superposition, and imperfections can always be projected on a
discrete set and thus can be corrected~\cite{Knill1997}. When such a
digital quantum simulation operates with fault-tolerant quantum
error correction~\cite{Preskill1997}, we can trust its output to a
known finite precision.

Universal digital quantum computers may serve as digital QSs (DQSs) that mimic  dynamics of some quantum
many-body system of interest. Despite the fact that in such a case
error correction and fault tolerance is guaranteed, the question
of efficiency of such a device is highly non-trivial. The number
of resources needed for precise simulation of continuous-time
dynamics of a many-body system by stroboscopic digital
applications of local gates might be enormous~\cite{Brown2006}. One can also
consider DQSs that are experimental systems
that have at their disposal only a limited, non-universal  set of
gates. In such a situation, the error correction and
fault tolerance are not guaranteed and the question of efficiency
is even more pertinent.

 This
raises the central problem of this key issue article: can we trust
the results obtained with a real-world {\em analog} or {\em
digital} QS, and under what conditions are they
{\em reliable} to a known degree of uncertainty? 
Although our main discussion concentrates on analog QSs (AQSs), the article reports also on the state of art of DQSs. It is organized as follows. Section II develops
the general concept of QSs in the spirit of the
DiVincenzo criteria for quantum computing~\cite{DiVincenzo2000}.
Here we present one of the main results of this article:  
a definition of the QS based on four
properties that a QS should have: relevance,
controllability, reliability, and efficiency.

Section III is devoted exclusively to DQSs. It contains several subsections in which we review various
proposals for DQSs, classify them, and
discuss the present state of knowledge concerning their
controllability, reliability, and efficiency. Section IV is
organized similarly,  but focused on 
AQSs. 

Section V is perhaps the most important one from the
conceptual point of view. Here, we formulate specific proposals to investigate the robustness of AQS and how to extend standard
methods of validation and certification of AQS. 
We illustrate these considerations in Sections VI-VIII 
with calculations for a paradigmatic model that are not published elsewhere. 
Section~VI describes the investigated model, and the results concerning statics and
dynamics are presented in Sections VII and VIII, respectively. 
We conclude in Section IX. The paper
contains also an Appendix describing  technical details of the
methods used.

Very recently, Nature Physics has published a focus issue with
five articles devoted to QSs: a short article by
J.I.~Cirac and P.~Zoller \cite{Cirac2012}, introducing the subject, and
four longer reviews on ultracold atoms \cite{Bloch2012}, ions \cite{Blatt2012}, CQED \cite{Houck2012}, and photons \cite{AspuruGuzik2012}. 
Our key issue article is complementary to these, in the sense that it
addresses general and universal problems of validation of quantum
simulations, their robustness, reliability, and efficiency -- problems that pertain to all kinds of QSs.

\section{Quantum Simulators}

Before proceeding, we must establish a clear definition of a
QS.  We consider here a QS to be a
device which, when measured, reveals features of an ideal
mathematical model, e.g., the phase diagram for the Bose-Hubbard
model on a specified lattice with specified interactions.  This
contrasts with, and is less demanding than, a full simulation of a
real material, since typically a mathematical model attempts to
capture only the most relevant properties of the real material.
For example, the superconducting properties of a cuprate might be
shared, in part, by a Fermi-Hubbard model~\cite{Lee2006,Lee2008}.  A
QS may be a special purpose device that simulates a
limited class of models, e.g, the Bose-Hubbard model simulated by
atom transport in an optical lattice~\cite{Jaksch1998, Greiner2001},
or a universal machine that is capable, in principle, of
simulating any Hamiltonian on a finite-dimensional Hilbert space.

Based on this, we formulate the following ``working'' definition of a quantum simulator in the
spirit of the DiVincenzo criteria for quantum
computing~\cite{DiVincenzo2000}, providing some more detailed explanations below (see also \cite{Lewenstein2012}):\\

\noindent {\bf Definition}\\

A QS is an experimental system that mimics a simple
model, or a family of simple models of condensed matter (or high-energy physics, or quantum chemistry, \dots). 
A quantum simulator
should fulfill the following four requirements:
\begin{itemize}

\item {\bf (a) Relevance:}  The simulated models
should be of some relevance for applications and/or our
understanding of challenges in the areas of physics mentioned
above. 

\item {\bf (b) Controllability:} A
QS should allow for broad control of the parameters
of the simulated model, and for control of preparation,
initialization,  manipulation, evolution, and detection of the
relevant observables  of the system.

\item {\bf (c) Reliability:} Within some prescribed error, one should be
assured that the observed physics of the QS
corresponds faithfully to that of the ideal model whose properties
we seek to understand.

\item {\bf (d) Efficiency:} The QS should
solve problems more efficiently than is practically possible on a
classical computer.

\end{itemize}

\begin{table*}
\begin{tabular}{  >{\flushleft}p{0.12\textwidth}     >{\flushleft}p{0.18\textwidth}  >{\flushleft}p{0.11\textwidth}  >{\flushleft}p{0.14\textwidth} >{\flushleft}p{0.02\textwidth} >{\flushleft}p{0.09\textwidth}  >{\flushleft}p{0.185\textwidth}  >{\flushleft}p{0.1\textwidth} }
  \toprule
			\multicolumn{1}{l }{} & \multicolumn{3}{ l }{{\bf {\large digital}}} 	& &	    \multicolumn{3}{  l }{{\bf {\large analog}}}		\tabularnewline
  \cmidrule{2-4} \cmidrule{6-8}
			\multicolumn{1}{l }{} & \multicolumn{1}{ l }{{\bf universal}}& \multicolumn{1}{ l}{{\bf non-univ.}}	& \multicolumn{1}{ l }{{\bf open}}& & \multicolumn{1}{ l}{{\bf universal}}& \multicolumn{1}{ l}{{\bf non-universal} }&	\multicolumn{1}{ l}{{\bf open}}	\tabularnewline
  \midrule
	{\bf Realizations}	& trapped ions, ultracold neutral  \newline or Rydberg atoms, circuit QED, super-\ conducting qubits, \dots	
			& same as \newline universal \newline digital	
			& as universal digital {\footnotesize (especially trapped ions, ultracold neutral or Rydberg atoms)}
			&
			& {\bf ?}
			& many  {\footnotesize (trapped ions, ultracold atoms, pho- tonic and polariton systems, artificial solid-state lattices, \dots )	}	
			& same as non-universal analog	\tabularnewline
	{\bf Control}		& full {\footnotesize (long-range inter- actions difficult ?)}	
			& partial	
			& partial	
			&
			& full 		
			& partial {\footnotesize (but long-range interactions ``easy'')}	
			& partial	\tabularnewline
	{\bf Error correction (EC)} & with exponential overhead {\footnotesize (Trotterization issues)}
			& not \newline guaranteed
			& not \newline guaranteed
			&
			& no \newline standard \newline EC
			& no \newline standard \newline EC 
			& no \newline standard \newline EC \tabularnewline
	{\bf Reliability}	& full 		
			& not \newline guaranteed
			& not \newline guaranteed 	
			&	
			& {\bf ?}
			& {\bf ?} {\footnotesize (partial \newline validation \newline schemes \newline available)}
			& {\bf ?} {\footnotesize (partial validation schemes available)}	\tabularnewline
	{\bf Efficiency} & efficient without EC {\footnotesize (for general class of models)}; much less efficient with EC {\footnotesize (Trotterization issues)}
			& at least as universal \newline digital, but \newline may not be provable
			& can be better than universal digital
			&
			& {\bf ?}
			& {\bf ?}		
			& {\bf ?} 		\tabularnewline
	\bottomrule
\end{tabular}
\caption{In this table, we characterize the different classes of quantum simulators (digital vs.\ analog, universal vs.\ non-universal, Hamiltonian vs.\ open) focussing especially on the requirements (b)~to~(d). (Since the relevance (a) depends on the concrete model simulated, we do not list it here.)
Note specifically that little is known about reliability and efficiency of AQSs (question marks). 
Detailed descriptions are provided in Sections~III and~IV.
\label{tab:QSs}}
\end{table*}

In Table~\ref{tab:QSs}, we summarize to which extent existing experimental proposals fulfill these requirements. We will characterize the different types of QSs in more detail in the following two sections, but before that, we would like to make some general comments. \\

\noindent{\bf Comments ad a)} We should demand that the mimicked models
are not purely of academic interest but that they rather describe some
interesting physical systems and solve open problems. 
This means also that the simulated models should be computationally very hard for classical computers (see also requirement (d)).\\

\noindent{\bf Comments ad b) and c)}  
Regarding control over measurable observables, one should stress that very
often the amount of output information required from quantum
simulators might be significantly smaller than one could
demand from a universal quantum computer. Quantum simulators
should provide information about phase diagrams, correlation
functions, order parameters, perhaps even 
critical exponents or nonlocal hidden order
parameters.  But a common assumption is that these quantities 
are more robust than what is required for a universal quantum computer, which typically relies on much higher-order correlation functions than a QS.

Regarding control over model parameters, it is in particular desirable to be able to set the parameters in a regime where the model becomes
tractable by classical simulations, because this provides an elementary instance of validating the QS. 
Furthermore, one of the  main results of this paper is the proposal and analysis of an even more sophisticated manner of validation, namely the checking of the sensitivity of the quantum simulation with respect to addition of noise and/or disorder. Such a validation is only possible with sufficient control over the system. 
Note, however, that there are other possibilities of checking the results, as pointed out to us by Z.~Hadzibabic \cite{HadzibabicPrivate}.
Namely, sometimes it is impossible to
simulate the system classically, but it might still be possible by classical means to test the sensibleness of the quantum-simulation results. For instance, the measured ground-state energy should fulfill all known bounds, such as variational ones, 
and others.\\

\noindent{\bf Comments ad d)} 
The notion ``computationally very hard for classical computers'' may have
several  meanings: i) an efficient 
(scalable, with polynomial growth in resources as a function of problem size) classical algorithm to simulate the model might not exist, or might not be
known; ii) the efficient scalable algorithm is known, but the required size
of the simulated model is too large to be simulated under
reasonable time and memory restrictions. The latter situation, in
fact, starts to occur with the classical simulations of the Bose-- or
Fermi--Hubbard models \footnote{In the temperature regimes of current experiments on the Fermi--Hubbard model, the model can be simulated classically by high-temperature expansion, but as soon as experiments achieve lower temperatures, this will cease to work and problems of fermionic simulations (i.e., sign problem in QMC \cite{Sandvik2010a}) will become relevant. On the other hand, recent efforts in variational Monte Carlo \cite{Sorellalecturenotes} and fermionic tensor networks \cite{Pineda2009,Corboz2009,Barthel2009,Shi2009,Corboz2010} are rapidly providing ever better variational approximations.}, in contrast to their experimental quantum
simulators. However, there might  be exceptions to the general rules.
For instance, it is desirable to realize QSs to
simulate and to observe novel phenomena that so far are only
theoretically predicted, even though it might be possible to
simulate these phenomena efficiently with present computers.
Simulating and actually observing in the lab is more than just 
simulating abstractly on a classical computer.\\

\noindent{\bf Comments ad c) and d)} The requirements of
reliability and efficiency are interrelated. In fact, we could try to
improve the precision of a QS by averaging more
experiments, but in hypersensitive regimes (like close to quantum phase transitions) the necessary number of
repetitions can grow rapidly, bringing the overall efficiency of
the QS down to the level of classical computers.  A
connection between (c) and (d) could also be relevant for the popular
cross-validation approach~\cite{Leibfried2010}. There, one compares the results of two different physical
realizations performing a QS of the same model, and hopes to find universal features which then would be ascribed to the simulated model. 
It may be, however, that the universal features shared by multiple platforms are robust only because they could have been predicted efficiently with some classical algorithm.\\

With our working definition at hand, one could ask --
What should a QS simulate? An important set of tasks include: 

\begin{enumerate}
\item Statics of the mimicked system at zero temperature; this implies
 ground-state simulation and its properties.

\item Statics at thermal equilibrium, i.e., Hamiltonian dynamics at low energies or
thermodynamics  at non-zero, typically low, temperatures.

\item Continuous-time dynamics of the system, in particular Hamiltonian dynamics
out of equilibrium.

\item Dissipative or open-system continuous-time dynamics.
\end{enumerate}

To understand which of these are most relevant, we now discuss shortly which systems
can be simulated efficiently classically and which systems are
classically computationally hard. Classical simulations of quantum systems are currently performed
using one of the following numerical methods \cite{Lewenstein2012}:
\begin{itemize}
\item Quantum Monte Carlo (QMC)
\item Systematic perturbation theory
\item Exact diagonalizations
\item Variational methods (mean field methods, density-functional theory (DFT),
dynamical mean-field  theory (DMFT), tensor-network states (TNS),
density-matrix renormalization group (DMRG), tree tensor network states (TTN), multiscale
entanglement-renormalization ansatz (MERA), projected entangled-pairs states (PEPS), ...)
\end{itemize}
Each of these methods has its limitations.
Let us first focus on points 1) and 2) of the previous list of possible QS tasks. In these cases, QMC works for various large systems, but fails for Fermi or frustrated systems due to the famous sign problem \cite{Sandvik2010a}.  Perturbation theory works only if there exists a small expansion parameter \cite{Dusuel2004}.  Exact diagonalization works only for rather small systems \cite{Sandvik2010a}.
In the case of 1D systems, DMRG, MERA and TTN techniques scale favorably and can, in principle, treat very large systems \cite{Schollwoeck2011, Alba2011b,Evenbly2011}.  In 2D the situation is more complex --  similar to exact diagonalization, DMRG and TTN work only for reasonably small systems \cite{Stoudenmire2011, Tagliacozzo2009,Tagliacozzo2011}, whereas  2D tensor-network methods (PEPS, MERA) in principle work for arbitrarily big systems (bosonic, and even fermionic \cite{Corboz2011} or frustrated \cite{Wang2011}) but are biased towards slightly entangled states.  Mean field \cite{Binney1992}, DFT \cite{Martin2004,Stoudenmire2011a}, or DMFT \cite{Georges1996}, finally, have other limitations, e.g., they are essentially designed for weakly-correlated systems. 

Which are then the models that are computationally hard for points 1) and 2) in the previous task list? 
Generally speaking, computationally hard are those ``strongly entangled'' models in more than 1D such as
\begin{itemize}
\item Fermionic models, with  paradigmatic examples being the
Fermi-Hubbard or $t-J$ models for spin 1/2 fermions \cite{Lee2006}.
\item Frustrated models, with  paradigmatic examples being
antiferromagnetic Heisenberg or $XY$ models on a kagom\'e or
anisotropic triangular lattice \cite{Misguich2004}.
\item Disordered models, with paradigmatic models being quantum, or even
classical spin glasses \cite{Das2008}.
\end{itemize}

When we move to points 3) and  4) of the task list, i.e., studying  dynamics, one can safely state that
\begin{itemize}
\item Quantum dynamics on a long time scale is generically
computationally hard.
\end{itemize}
The latter statement implies that while it might be possible to
simulate with classical computers short-time dynamics in a restricted class of 1D models, such attempts  will nearly
always fail at longer time scales. Indeed, this fact is related to
correlation and entanglement spreading according to the Lieb--Robinson
theorem that states that, after a sufficiently large time, states can become strongly entangled
(\cite{Lieb1972,Bravyi2006,Calabrese2006,Eisert2006,Nachtergale2006,Schuch2008a},
see also \cite{Eisert2010}).

In the following two sections, we will explore in more detail the state of the art concerning the four requirements (a-d) of our definition, first for digital, then for analog QSs.

\section{Digital Quantum Simulators (DQS)}

In this section, we  classify DQSs, discuss their general properties, various protocols for implementing such devices, and summarize state-of-art knowledge concerning their controllability,
reliability, and efficiency.

\subsection{Universal Digital Quantum Simulators (UDQS)}

While the concept of QSs should be traced back to
prophecies of Feynman  \cite{Feynman1982}, the ideas were made concrete by Lloyd who showed that any ``local'' many-body unitary evolution governed by a ``local'' Hamiltonian could be implemented by the control afforded by a universal digital quantum computer \cite{Lloyd1996}.  
For this reason, in the following we will term  Lloyd's DQS a ``universal DQS'' (UDQS).

Lloyd's UDQS is in fact a universal quantum computer, whose task is
to simulate the unitary time-evolution operator of a
certain quantum system described by a physical Hamiltonian, 
which can then be employed to extract quanties like energy gaps and ground-state properties.   
This is done by appropriate subsequent stroboscopic applications of
various quantum gates that mimic the action of a global
unitary continuous time evolution operator of the system.
The mathematical basis for such a digitalization is given by the
Trotter--Suzuki formula. In order to realize Lloyd's UDQS in a laboratory,  the experimentalist has to have to his/her disposal a universal set of unitary quantum gates 
\footnote{
Some authors (cf.\ \cite{Cirac2012,Blatt2012}) use the term UDQS for DQS that can simulate all possible evolutions governed by  arbitrary (or, more precisely, arbitrary ``interesting'') Hamiltonians. We put the emphasis here on the unversality of the set of gates, rather than the set of simulable Hamiltonians. 
}.
 Let us list below some possible realizations and properties of UDQSs:
\begin{itemize}

\item {\bf Realizations:} While implementation of a fully-functioning large-scale digital quantum computer is still in
 development, there are several physical systems for which the universal
 sets of quantum gates are available, and for which realization of
proof-of-principle UDQSs is possible. These systems include
ultracold ions \cite{Blatt2012}, ultracold trapped
atoms interacting via cold collisions \cite{Bloch2012}, or the Rydberg-blockade mechanism \cite{Mueller2009,Weimer2010},
circuit QED \cite{Houck2012}, superconducting qubits
(for reviews see \cite{Makhlin2001,Devoret2004,You2005,Girvin2009}, for the current
state of art \cite{Reed2012} and references therein). For a general overview see
also \cite{Zoller2005}. The first concrete proposals for
realization of UDQSs where given in ~\cite{Jane2003,Wiebe2011},
and the first experiments, perhaps, 
were performed in NMR systems \cite{Somaroo1999,Tseng1999,Aspuru-Guzik2005,Du2010}. Using a digital architecture
and stroboscopic sequence of gates, the quantum simulation of 
Ising, XY, and XYZ spin models 
in a transverse field were recently demonstrated in a 
proof-of-principle experiment with up to six ions~\cite{Lanyon2011}.

\item  {\bf Controllability:} In accordance with \cite{Lloyd1996}, a UDQS is perfectly
controllable, i.e., with the help of a universal set of gates sufficient
control of the parameters can be achieved. This control  allows
for simulation of practically any local Hamiltonian evolution, as
well as for preparation, manipulation, and detection of relevant
states and observables of the system in question. 
Further, Preskill's group has
proven recently that the scattering amplitudes in the simple
relativistic quantum field theories can be efficiently (in
polynomial time) simulated by UDQSs \cite{Jordan2011,Jordan2011a}.
Note, however, that not much is known about the possibility of quantum simulation of
systems with long-range interactions like Coulomb or dipole--dipole
interactions using UDQSs. 

\item {\bf Error correction:} A UDQS is the only DQS which has guaranteed access to error
correction and fault tolerance \cite{Shor1995,Nielsen2000} (for the first proof-of-principle experiments see
\cite{Cory1998,Chiaverini2004,Pittman2005,Lassen2010,Schindler2011,Reed2012}).

\item{\bf Efficiency:}  So far, the community has mostly focused on developing requirement (b) for suitable relevant models,
 both theoretically and experimentally. The conditions (c) and (d) have received considerably less
attention, especially their interrelation.  Most work is focused
on efficiency in the absence of errors.  Lloyd showed that a
Trotter--Suzuki decomposition of a time-evolution operator is
efficient in that each logic gate acts on a scalable Hilbert space
associated with a small subset of qubits and the total number of
gates $N$ scales polynomially, $N\sim t^2/\epsilon$, where $t$ is
the time of evolution to be simulated and $\epsilon$ is the error
in the result \cite{Lloyd1996}.  Aharanov and Ta-Shma showed that a UDQS is efficient when the Hamiltonian is ``sparse,'' i.e., the number of nonzero entries in any row is at most poly(log($D$)), where $D$ is the dimension of the many-body Hilbert space~\cite{Aharonov2003}.  In the absence of errors, the computational complexity of such a simulation has been well studied~\cite{Berry2007, Childs2011}.

\item{\bf Reliability:} In the presence of errors, however, ensuring reliability to a
desired precision has profound implications for efficiency even in a digital simulator on a
fault-tolerant quantum computer~\cite{Brown2006, Brown2010}. In the digital approach with a finite universal gate set, one applies
error-correction schemes that can make the whole computation
fault-tolerant when the error per operation is below a certain
threshold -- thus digital simulators fulfill the reliability
requirement (c).   The Trotter expansion, however, can scale poorly
when error correction is included, as emphasized by Brown {\em et
al.}~\cite{Brown2006}
in studies of an implementation of a quantum algorithm to calculate the low-lying energy gap in pairing Hamiltonians~\cite{Wu2002}. 
Because the number of gates in the
expansion scales as $1/\epsilon$, in order to achieve $M =
-\log_2(\epsilon)$ bits of precision, we must Trotterize the
unitary evolution to be simulated into $2^M$ slices.  In the
presence of errors,  each of the time slices must be implemented with only a finite set of universal gates, according to the Solovay-Kitaev theorem~\cite{Kitaev2002}; only then can they be implemented fault tolerantly.  The result is that a fault-tolerant implementation of the Trotter
expansion requires a number of
gates and time to perform the simulation that grows {\em exponentially} with the degree of precision required, for a fixed number of particles being simulated.  
Moreover,  Brown {\em et al.} showed that for a small number of qubits where one
might avoid error correction, analog control errors on the logic
gates can lead to faulty results, negating requirement (c), and
robust control pulses become essential. 

In a similar vein, Clark {\em et al.}~\cite{Clark2009} 
performed a careful analysis of the resources necessary to
implement  the Abrams-Lloyd algorithm~\cite{Abrams1999} to calculate
the ground-state energy of the one dimensional transverse Ising model (TIM) using a state-of-the-art fault-tolerant architecture for an ion-trap quantum computer.  
Again, the overhead in the number of time steps to fault-tolerantly implement the quantum-phase estimation algorithm grows exponentially with the degree of precision required. 
They found that for 100 spins, in order to achieve $b\ge 10$ bits
of precision, at least two levels of concatenated error correction are necessary, requiring at least 100 days of run time on the ion-trap quantum computer; for $b\ge 18$, three levels are
necessary, requiring at least $7.5 \times 10^{3}$ years!  
These results assume a gate time of $10\mu\mathrm{s}$. To recover the 100 days limit with only one level of concatenated error-correction coding, a gate time of $300\mathrm{ns}$ seems necessary (as well as decreasing other parameters such as failure probabilities). 
On the other hand, for a fixed precision, the number of resources required grows weakly with system size.  So, if the error probability per gate can be reduced well below threshold to achieve the desired precision without many layers of concatenated error-correction encoding, then digital quantum simulation will scale favorably with the number of particles.
\end{itemize}

Let us finally remark that, to increase their efficiency, digital-quantum-simulation algorithms
often compress the number of degrees of freedom that are necessary
to describe the many-body system, rather than directly map the
Hilbert space of the system to the Hilbert space of the
simulator~\cite{Brown2010}, an approach that has been borrowed from classical algorithms like MPS or PEPS. 
Currently, there is a new theoretical development towards a ``hybrid'' device,
where the ground state of many-body Hamiltonians is represented as
a PEPS, but implemented on a quantum computer. This is efficiently possible when the gap between the ground and first excited state scales as
the inverse of a polynomial in the number of particles.  
Then, one can use the quantum computer to contract tensor networks and use that to calculate the expectation
value of any local variable, such as correlation functions~\cite{Schwarz2011}.   
In a similar spirit, Temme {\em et al.} developed a quantum-algorithmic version of the Metropolis Monte-Carlo algorithm that allows one to efficiently sample from a Gibbs thermal
state~\cite{Temme2011}. 
Such approaches point to efficient DQSs for well defined classes of problems.

\subsection{Non-Universal Digital Quantum Simulators 
(nUDQS)} 

A non-Universal Digital Quantum Simulator (nUDQS) is in many aspects
similar to a UDQS, except that is it a special-purpose quantum
computer. Its task is, however, the same as that of a UDQS: to
simulate continuous-time quantum many-body dynamics of a
certain quantum system described by a certain  physical
Hamiltonian.   The experimentalist who realizes a nUDQS has to
his/her disposal a non-universal set of unitary quantum gates. Let
us list below some properties and possible realizations of such a nUDQS:
\begin{itemize}

\item {\bf Realizations:} In all systems in which the universal
 sets of quantum gates are available, one can also restrict the set of gates and realize a nUDSQ.
For example, in some of the recent experiments of Blatt \cite{Lanyon2011}, only a necessary subset of the available set of universal gates was used. 
All of the systems discussed above (atomic, superconducting, etc.)\ are potentially platforms for implementing nUDQSs.  A seminal example of this approach goes back to 
the so-called  ``Average Hamiltonian Theory'' in nuclear magnetic
resonance (NMR) \cite{Haeberlen1968,Waugh2007}.

\item {\bf Controllability:}  nUDQSs are typically not  perfectly
controllable, but in most experimental realizations should allow for a
wide control of parameters, which in  turn should  allow for
simulations of evolution for wide families of  Hamiltonians of
interest.

\item{\bf Error correction:} 
For nUDQSs, it is not guaranteed that error correction and fault-tolerant computing is possible.

\item{\bf Efficiency and reliability:} All of the above discussion concerning UDQSs
applies also to nUDQSs. 
But, there are many novel, open problems associated specifically with nUDQSs, since, e.g., sometimes giving up on universality can result in substantial efficiency gains. 
For example, universality could be sacrificed in favor of a highly precise and fast gate \cite{Lanyon2011} (a simple example is an external homogeneous field, which in a UDQS might have to be applied as a sequence of one qubit gates).
In particular, it is possible that for some classes of nUDQSs the problems of Trotterization are not as severe as in the case of UDQSs \cite{Weimer2010}.

\end{itemize}

\subsection{Open-System Digital Quantum Simulators 
(OSDQS)}

An Open-System  Digital Quantum Simulator (OSDQS) is a completely new
concept, in principle very different from DQSs aimed at Hamiltonian
evolutions. OSDQSs are designed to simulate open-system,
dissipative dynamics described in the simplest situation by a
Markovian Lindblad master equation for the density matrix of
a many-body system of interest. OSDQSs can be aimed at a continuous-time simulation of interesting open-system dynamics, or at a designed dissipative dynamics toward a stationary state of
interest, in particular a pure, highly-entangled state \cite{Diehl2008,Kraus2008,Verstraete2009}.

The experimentalist who realizes an OSDQS, in contrast to a UDQS or a nUDQS,  needs to
have at his/her disposal some non-unitary, dissipative quantum gates, which mathematically correspond to Lindblad super-operators
acting on the density matrix in the master equation.  This fact
opens a plethora of new questions, e.g., what are the universal sets of gates for this type of evolution.
Note that in the case of unitary computing, the universal set of gates allows for realization of
arbitrary unitary transformations acting on the (pure) state of the
system. In the case of open-system dynamics, a universal set of gates
should allow for the realization of an arbitrary 
completely positive map (CPM) acting on the density matrix of a system. Moreover, for
experimental realizations, we require the gates to be local. 

While the conditions for controllability of an open quantum systems are under
exploration~\cite{Wu2007a}, the question of a universal set of gates
in this context remains open. A non-trivial reduction (cf.\
\cite{Wolf2008a,Cubitt2009}) of this question to the CPMs that
correspond to Markovian evolution, is also open. 
The problem of error correction in this context is unsolved as well. 
All of these comments imply that in the area of OSDQSs there are more open
questions than answers.

\begin{itemize}

\item {\bf Realizations:} 
In systems in which the universal set of quantum gates is available, one way to realize dissipative gates is by tracing out ancillas, thus allowing to realize an OSDQS.
Good testbeds for exploring OSDQSs are provided by Rydberg atoms, atomic
ensembles, 
NMR \cite{Cucchietti2010}, or trapped ions. In fact, the first concrete proposals
for open system DQS concerned Rydberg gates
\cite{Mueller2009,Weimer2010}. The first experimental realizations of
these ideas, however, have been achieved with
trapped ions \cite{Barreiro2011}.

\item {\bf Controllability:}  OSDQSs are typically not  universal
since they are not usually controllable in the sense of realizing
an arbitrary quantum map \footnote{Often we do not even know if
they are controllable: to our knowledge the problem what is the
universal set of non-unitary ``gates" that allows to realize an
arbitrary, say, completely positive linear map for an $N$-qubit
density matrix has not been directly solved.}. Nevertheless, many
experimental realizations should allow for a wide control of
parameters, which in  turn should allow for simulations of
open-system (Markovian) evolutions for wide families of open
systems of interest. As pointed out in
\cite{Kraus2008, Verstraete2009}, due to the
purely dissipative nature of the process, this way of doing
quantum information processing exhibits some inherent robustness and defies
some of the DiVincenzo criteria for quantum computation. In
particular, there is a natural class of problems that can be solved
by open-system DQSs or AQSs: the preparation of ground states of
frustration-free quantum Hamiltonians.

\item{\bf Error correction:} For OSDQSs, it is not guaranteed that error
correction and fault-tolerant computing is possible in the sense
defined above \footnote{The standard schemes for error correction
assume that the quantum computer (i.e., DQS) follows a unitary
evolution, i.e., dissipation and decoherence are considered there
as sources of errors, which the error correction is supposed to
remove.  To our knowledge, there are no works where these are
considered as desired, although not perfect, which error
correction is supposed to restore. In this sense, since there are
no known ways for error correction under these circumstances, error correction is ``not guaranteed,'' as we write above.}

\item{\bf Efficiency and reliability:} All of the above discussion concerning UDQSs
and nUDQSs applies also to OSDQSs. 
But, due to  purely
dissipative nature of the process, this type of simulation has a
certain intrinsic robustness and built-in ``error correction.'' A clear example is seen in the OSDQS implementation of Kitaev's toric code \cite{Weimer2010,Barreiro2011}. 
However, as discussed in Ref.~\cite{Weimer2010}, errors in the gates result in effective heating. 
Also, the problems of
Trotterization are not as severe as in the case of quantum
simulators of Hamiltonian evolution. Still, most of these general
aspect concerning OSDQSs have not yet been investigated
systematically. The efficiency of OSDQSs for the case of
frustration-free Hamiltonians depends on the size of the gap between the ground state and the excited states, or more precisely on the real part of the first non-zero eigenvalue of the Lindblad equation, which determines the rate of approaching the stationary (ground) state.

\end{itemize}

Currently, considerable attention has been devoted to the
problem of existence and uniqueness of the open-system
preparation of ground states of frustration-free Hamiltonians, and
in particular, entangled states of interest. These states are
annihilated simultaneously by all of the local frustration-free
Lindblad superoperators entering the master equation. There is little 
known in general about the many-body dissipative dynamics with a 
frustrated set of Lindblad superoperators competing with
Hamiltonian
dynamics. For the first attempts to understand these
kind of problems in the context of quantum diffusion-exclusion
processes competing with Hamiltonian evolution see, e.g., 
\cite{Temme2009}.

\section{Analog Quantum Simulators 
(AQS)}

In this section, we classify general properties of AQSs, discuss various proposals for such devices, and 
summarize the state-of-art knowledge
concerning their controllability, reliability, and efficiency. AQSs
are experimental systems that are designed to mimic the quantum
dynamics of interesting quantum many-body models, typically using 
``always on'' interactions between particles that are augmented by fast local unitary control.
While by definition they operate in continuous time and thus the
Trotterization problems do not concern them, the standard error-correction methods  and fault tolerance cannot be applied.

\subsection{Universal Analog Quantum Simulators 
(UAQS)}

Sometimes known as ``Hamiltonian simulation,'' the goal of a UAQS is to transform a given Hamiltonian acting on a fixed Hilbert space into an arbitrary target Hamiltonian through a well-designed control sequence. While not conceived as a practical AQS device, the protocol explores an abstract quantum-information-processing system capable of simulating unitary evolution for  all (or at least all local) Hamiltonians.

\begin{itemize}

\item {\bf Realizations:} To our knowledge there are no concrete
proposals for experimental realizations of UAQSs.

\item  {\bf Controllability:}  
While for UDQSs the issue is the access to the universal set of quantum gates, for UAQSs the
question is what the necessary resources are (not necessarily quantum
gates) that allow for the simulation of all Hamiltonian  evolutions
of interest. Universal  control sets (as opposed to universal digital logic gates) that generate an arbitrary Hamiltonian evolution have been studied~\cite{Dodd2002,Wocjan2002}.  Typically, such an approach using ``always on'' interactions is associated with more limited control than is available in a universal digital quantum computer.

\item {\bf Error correction:}  UAQSs do not allow for standard error correction and
fault tolerance.

\item{\bf Efficiency and Reliability:}  
D\"{u}r {\em et al.} studied a hybrid construction of always-on interactions with stroboscopic digital control to achieve a universal Hamiltonian simulator via the Trotter construction~\cite{Duer2008}.  They found that decoherence and analog timing errors can make this inefficient for a Hamiltonian simulator.  Other issues concerning UAQSs are essentially the same as for non-universal AQSs, so we leave the discussion of them to the next subsection. The only difference
is that UAQSs, by definition, are capable of performing tests of robustness of the quantum simulations that we propose in the following section, i.e., tests based on adding disorder or noise in a controlled manner to the simulated Hamiltonian. For non-universal AQSs such an addition requires additional resources.

\end{itemize}

\subsection{Non-Universal Analog Quantum Simulators (nUAQS), or simply AQS}

Non-universal AQSs constitute the most popular class of quantum
simulators, but despite this fact, there is very little known about their reliability and efficiency. 
Therefore, we focus on them in the remainder of this paper, where, for simplicity, we shall term them AQSs. 
AQSs are experimental systems that can mimic
continuous-time unitary Hamiltonian evolution  for a family of
models of many-body physics. Their characteristics are as follows:

\begin{itemize}

\item {\bf Realizations:} The most advanced experiments with  AQSs
have been with ultracold atoms in optical lattices 
\cite{Lewenstein2012,Bloch2012,Greiner2001,Lin2009b,Bakr2010,Trotzky2010,Jordens2010,Baumann2010,Liao2010,Simon2011,Struck2011a,VanHoucke2012,Trotzky2012,Cheneau2012}.
The degree of quantum control is even better in ultracold-ion
systems, but these are so far limited to few ions
\cite{Friedenauer2008,Kim2010,Islam2011,Gerritsma2010,Gerritsma2011}.
The first step toward large-scale QSs with ions was, however, recently achieved
\cite{Britton2012}. 
Recently there has also been substantial progress in investigations of
other possible candidates for AQS, such as photonic systems
\cite{Lanyon2007,Lanyon2008,Belgiorno2010,Kitagawa2011,Ma2011},
photonic and polariton systems
 \cite{Greentree2006,Hartmann2006,Angelakis2007,Wang2009a,Koch2010,Lanyon2010},
artificial lattices in solid state systems \cite{Singha2011}.

\item  {\bf Controllability:} Most, if not all  of the proposals for and realizations of
AQSs allow for at least partial controllability.
The paradigm examples are AQSs employing ultracold atoms in optical
lattices (for more details, see Chapter 4 of \cite{Lewenstein2012}).
Here, the typical controls involve optical lattice parameters
(laser intensity, wavelength, etc.), lattice geometry, lattice
dimensionality, temperature and other thermodynamical control
parameters, as well as atomic interaction strength and nature (van der Waals
interactions are controlled via Feshbach resonances, while dipole
interactions by the strength of the dipoles, lattice-site-potential
shape, etc.). Further, tunneling can be laser assisted and can  mimic
artificial Abelian or even non-Abelian gauge fields (cf.\ \cite{Aidelsburger2011,Jimenez2012}).
Dipolar interactions may lead to non-standard terms in Hubbard
models, such as occupation dependent tunneling~\cite{Sowinski2012}, or
various effects involving higher orbitals (see, e.g.,~\cite{Dutta2011}). 

\item {\bf Error correction:} AQSs do not allow for standard error correction and fault
tolerance.

\item{\bf Efficiency:} The issues of reliability and efficiency are  essential for
the usefulness of any QS, and AQSs in particular. 
In the context of AQSs, however, there has been little
analysis of these problems. Firm criteria on computational complexity and efficiency for AQSs are in general
difficult to address and have not yet been established. First of
all, they require the knowledge of classical computational
complexity of the static or dynamical properties of the considered
quantum models. Unfortunately, in the realm of classical
computation, there are few proofs that a given computational
problem is outside the class $P$, or even if there is a clear
delineation between certain complexity classes. In recent years, there 
has been considerable progress in understanding that the
ground states of  1D gapped systems can be efficiently simulated
by classical methods \cite{Schuch2008b,Schuch2008c,Hastings2009}, or that
the quantum dynamics is in general computationally hard
\cite{Schuch2008a,Eisert2010}. If we can set the parameters of
our AQS to a regime where efficient classical simulation is
possible, we can assess the efficiency and reliability of the AQS in
this case by direct comparison with classical simulations (see below). However, there is no guarantee that such a calibration will hold in
the truly interesting regimes of parameters,  where efficient
classical simulations are either impossible, or we do not know how
to perform them.

\item{\bf Reliability:} So far,  there exists no perfect and rigorous way to assess the reliability of AQSs, but there are several
complementary approaches. One approach is by cross validation of a variety
of different physical systems (e.g., atoms in optical lattices,
ions in traps, and superconductors)~\cite{Leibfried2010}. The hope is that since every
platform has its own set of imperfections, they will agree on the
universal properties of the ideal quantum many-body model being
simulated. While it remains to be seen whether such universal
features would emerge, this approach has a number of shortcomings.
For example, there may be models that have only one known
implementation, or different implementations may suffer in the
same way from imperfections, hence consistently exhibiting
features associated with noise rather than with the ideal model.

A more systematic approach is to validate results of a quantum
simulator against analytical and numerical predictions in the
regime of parameters where such comparison is possible.  This was
recently demonstrated in experiments with ultracold bosonic and
fermionic atoms~\cite{Bakr2010,Trotzky2010,Jordens2010}; amazingly, in
one case numerical simulations helped to correct the expected
experimental temperature by up to $30\%$. Relying solely on
validation from classical calculations, however, would restrict
QSs to models in regimes where these efficient
classical algorithms exist -- that means  contradicting our
relevance and usefulness requirement, point~(a) of our
definition of a QS. In general, we desire to operate quantum
simulators in regimes whose properties are difficult to deduce by
classical methods, e.g., near or at the critical point of a QPT,
or in genuine {\it terra incognita} regimes. In these regimes, however, 
many relevant models become {\em hypersensitive} to
perturbations~\cite{Quan2006,Zanardi2006}, and even
small levels of noise may spoil completely the results of the
quantum simulation.  Indeed, the capability of an analog quantum
information processor whose dynamics is characterized by quantum
chaos (i.e., well described by random matrix theory) can be
severely impacted by imperfections~\cite{Georgeot2000, Frahm2004}.
More importantly, this also means that successfully validating a
QS in a classically accessible regime does not give
certainty about its robustness in regimes which are classically
not accessible.

\end{itemize}

\subsection{Open-System Analog Quantum Simulators 
(OSAQS)}

Finally, let us mention Open-System Analog  Quantum Simulators
(OSAQSs).  Similar to OSDQSs, OSAQSs are supposed to simulate
dissipative dynamics described in the simplest situation by a
continuous-time Markovian Lindblad master equation for the
density matrix of a many-body system of interest. OSAQSs can be
aimed at a  simulation of interesting open-system dynamics, or at
a designed dissipative dynamics toward a stationary state of
interest. 
Many-body Lindblad master equations have been studied in the context of evaporative \cite{Gardiner1997,Gardiner1997a,Jaksch1997,Gardiner1998,Jaksch1998a,Gardiner1998a,Gardiner2000a}, laser
\cite{Cirac1994,Cirac1994a,Cirac1995a, Morigi1997,Santos1999,Santos1999a,Castin1998,Idziaszek2001,Idziaszek2003,Idziaszek2003a,Dziarmaga2005,Dziarmaga2005a} and sympathetic \cite{Lewenstein1995,Mosk2001,Papenbrock2002,Papenbrock2002a,Wang2002} cooling of
degenerate atomic gases (see also \cite{Augusiak2010}). Recently, there
has been a revival of interest in such systems in the context of
possibility of using them for preparation of interesting pure,
highly-entangled states \cite{Diehl2008,Kraus2008,Verstraete2009}.
Open-system quantum simulators employing superconducting qubits may also give insight into exciton transport in photosynthetic complexes \cite{Mostame2011}.

The experimentalist who realizes an OSAQS, in contrast to an AQS  has to
have to his/her disposal some non-unitary, dissipative  quantum
mechanism: in a sense, all designed cooling  or entropy-reduction
methods are of this sort.

\begin{itemize}

\item {\bf Realizations:} All AQS systems can, in principle, be used as OSAQSs.

\item {\bf Controllability:}  OSAQSs are typically not  universal in a
sense similar to OSDQSs;  they allow neither for simulating
arbitrary (local) Markovian dynamics, nor do they allow for
preparation of arbitrary states. Nevertheless,  in most of the
proposals \cite{Diehl2008,Kraus2008,Verstraete2009} or
experimental realizations they allow for a wide control of
parameters, which in  turn allows for simulation of open-system
(Markovian) evolutions for wide families of open systems.

\item{\bf Error correction:} For OSAQSs, it is not guaranteed that error
correction and fault-tolerant computing is possible in  the sense
defined in the subsection on OSDQSs.

\item{\bf Efficiency and reliability:} All of the above discussion concerning AQSs applies also OSAQSs.
But, again due to the purely dissipative nature of the process, this
type of simulation has a certain amount of intrinsic robustness and built-in
``error correction'' - this is particularly clear for the
OSAQS of  quantum kinetic Ising models or Kitaev's toric code
\cite{Augusiak2010}.  Still, as in the case  of OSDQSs, most of
these general aspects concerning OSAQSs have not yet been
investigated systematically.

\end{itemize}

\section{Robustness of Analog Quantum Simulators}

All of the above considerations clearly lead to the fundamental
question: Can we trust quantum simulators? From what we have said,
the rigorous answer to this question is ``no'', yet in practice we
do tend to trust them, at least to some extent.

In order to gain more trust in the results of quantum simulations,
it is thus extremely important to design novel tests and
certification of reliability and validity of QSs. In this section,
which constitutes some of the most important results of this
paper, we propose such tests, which we call tests of
\textit{robustness of quantum simulators}. Our tests consist in
checking the robustness of QSs with respect to addition of
imperfections, such as static disorder or dynamical noise. 
This would then allow to (i) judge how strong the reaction of the QS with respect to these perturbations is, and (ii) might even open possibilities to extrapolate interesting observables to the ideal, zero-disorder limit. Such tests can also be applied to DQSs, but are particularly suited for AQSs.
For example, in an implementation with trapped ultracold atoms, disorder can be increased in a
controlled manner~\cite{Sanchez-Palencia2010}.

In the following of this key issue article, we use the example
of the quantum Ising chain to substantiate our discussion of the
reliability of AQSs and the relationship to the complexity/efficiency of the simulation. 
We study how imperfections affect the results of a AQS simulating that model, where, for simplicity, we assume quenched disorder as the only possible imperfection. 
In the future, it will be in particular interesting to also investigate the effects of dynamical noise, and the decoherence and relaxation that occurs due to coupling with an environment 
(see also Ref.~\cite{Zuniga-Hansen2012}). 

The quantum Ising model, which we describe in detail in Section VI, is exactly solvable, which allows us to explore regions with universal behavior such as second-order QPTs. 
In Section VII, we show that the ground-state expectation values of certain local observables appear fairly robust under disorder, while this need not be true for the global many-body state of the simulator. In particular, disorder can have a significant effect on relevant quantities that one could hope to extract from the simulator, such as critical points and exponents, or -- if the system is described by a conformal field theory (CFT)
-- its central charge~\cite{bookCardy}. Finally, we briefly address the relationship between robustness and complexity by studying the dynamics of different thermal states after a quench of the Hamiltonian (Section VIII). We show evidence that QSs appear to work better in regimes that are classically easier to solve or simulate (high-temperature states), thus hinting at a connection between the amount of quantum correlations and the robustness of a QS. \\

\section{The model}

To illustrate the influence of disorder on an AQS, we study the transverse
Ising model (TIM)
\begin{equation}
H=-\sum_{\langle i,j \rangle} J_{ij} \sigma_i^x \sigma_j^x
  -\sum_{i} h_{i} \sigma_i^z\,,
\end{equation}
where $\sigma_i^{x,z}$ are the usual Pauli spin matrices and $\sum_{\langle i,j \rangle}$
means sum over nearest neighbors.
The system is subject to quenched disorder in both the interaction and field terms.
We denote the nearest-neighbor spin coupling and the
transverse field by $J_{ij}=J(1+r\delta_{ij})$ and
$h_{i}=h(1+r\eta_{i})$, respectively, where $\delta_{ij}$ and
$\eta_i$ are independent random variables with a Gaussian distribution of mean
zero and variance $r$.
All details of our calculations are presented in the Appendix.

The TIM, even under the presence of disorder, is efficiently solvable -- by which we mean that the eigenstates and eigenenergies of the system
can be found using a classical computer, and that the cost of the algorithms (in time and hardware) is polynomial with the size (number
of particles) of the system. The TIM, in particular, can be solved
by using a Jordan--Wigner transformation to a system of non-interacting fermions 
-- and the cost of solving the non-interacting fermion system is the cost of diagonalizing a matrix with rank equal to twice the number of spins in the chain
\cite{deMartino2005}. This model is well studied (see for instance
\cite{Lieb1961}): for low fields, the ground state is a
ferromagnet, while for large fields it is a paramagnet. At zero
temperature and disorder, the system undergoes a QPT when the dimensionless control parameter
$\lambda=h/J$ approaches the critical value, $\lambda_c=1$, i.e.,
when the field intensity equals the interaction strength.
The influence of disorder can have dramatic effects on this phase diagram:
imperfections can create new phases, or even destroy the ones we want to investigate.
Indeed, in the TIM when the disorder strength is comparable with the interactions,
the critical point disappears and is replaced by a so called Griffiths phase~\cite{Griffiths1969},
extending across a region of size proportional to
the disorder strength.
Even more, in this Griffiths phase observables become non-self-averaging, i.e., fluctuations increase with system size, and hence dominate the thermodynamic limit.
In this study, we consider small disorder strengths, which allows us to ignore the Griffiths phase, especially in finite-size systems. Moreover, a state-of-the-art AQS can achieve very low levels of disorder, whence this is the experimentally relevant regime.
Note that, while the TIM has been studied extensively in the limit of very large disorder~\cite{Refael2004}, there are few studies addressing directly the influence of small disorder on the universal properties near quantum phase transitions. However, it is well known that even small disorder can lead to novel quantum phases such as Bose \cite{Fisher1989} or Fermi \cite{Freedman1977} glasses  (see also the section on disorder in Ref.~\cite{Lewenstein2007} and references therein).
In the following, we analyze the robustness of relevant observables to disorder in static and dynamic situations.

\section{Results -- Statics} 
\begin{figure}
\centering
\includegraphics*[width=8.6cm]{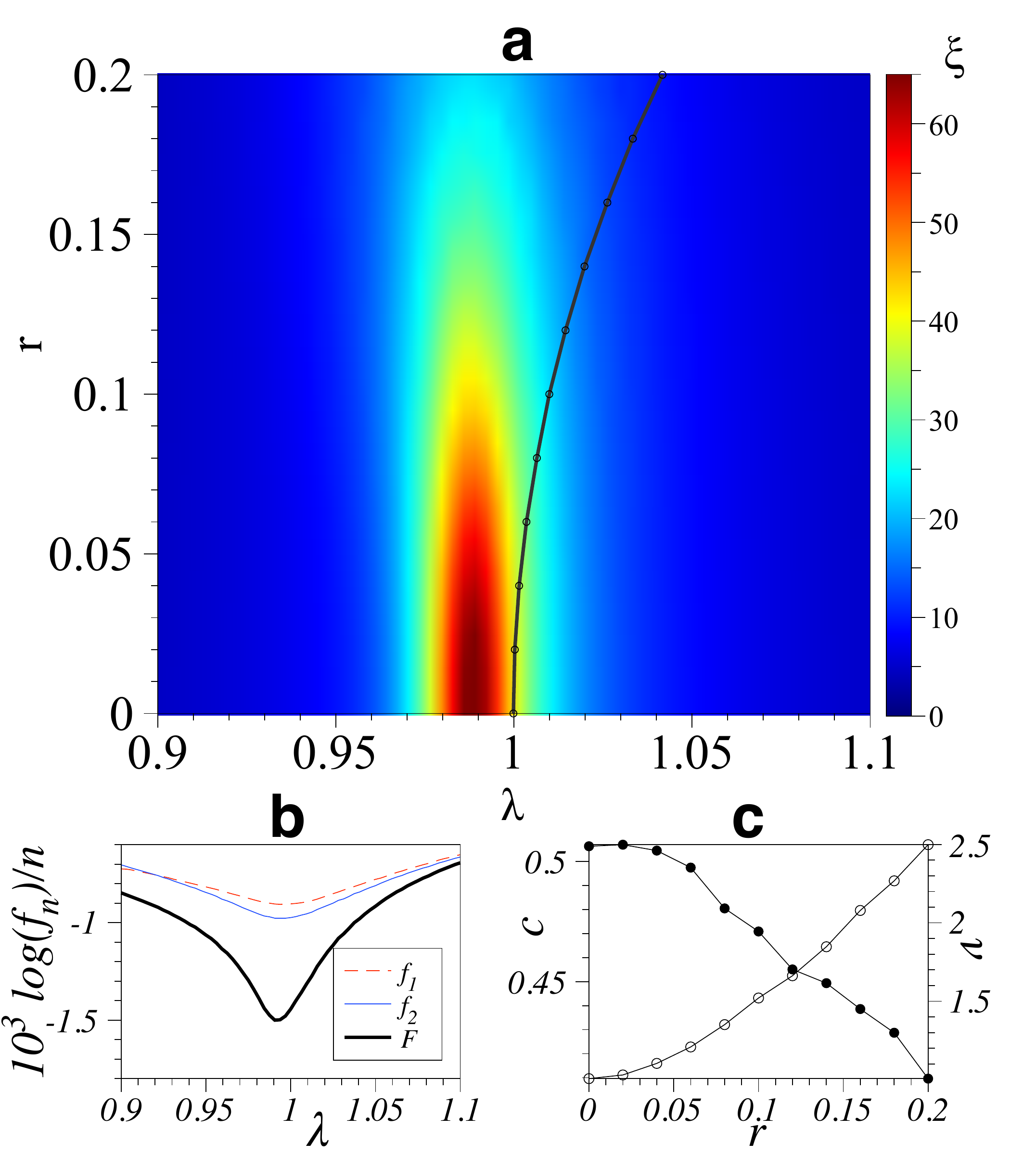}
\vskip-0.4cm
\caption{
 \textbf{A:} The correlation length $\xi$
decreases with disorder $r$, and its peak broadens (shown for a chain of 400 spins). 
The critical point (as extracted from a finite-size scaling of the energy gap $\Delta$) moves to larger $\lambda$ with increasing disorder (black line).
\textbf{B:}
For a chain of 400 spins, we show the mean single- and two-site reduced simulator fidelities ($f_1$ and $f_2$), and the total simulator fidelity ($F$) for a fixed disorder level $r=0.1$.
Local fidelities are more robust, which gives hope that local quantities can be
reliable even if disorder deteriorates the overall ground state.
As expected, disorder has more severe effects close to the QPT.
\textbf{C:}
The central charge $c$ (full circles, left axis), extracted from a fit to the part-chain entropy, and the critical exponent $\nu$ (open circles, right axis), extracted from a collapse of the correlations in different chain lengths.
Both change with disorder, which can lead to erroneously assigning the QPT to an incorrect universality class.
However, the change begins relatively smoothly at low levels of disorder.
} \label{FigureStatics}
\end{figure}

First, we investigate static
properties of the AQS and their robustness to
disorder (summarized in Fig.~\ref{FigureStatics}). We average all
the analyzed static quantities over many realizations of disorder.

One can evaluate the response of the AQS to disorder using the \textit{simulator fidelity},
which we define for pure states as the overlap between the state obtained with a perturbed simulator, $\ket{\Psi_r(\lambda)}$, and the ideal state $\ket{\Psi_0(\lambda)}$,
\begin{equation}
F(r,\lambda)=|\braket{\Psi_0(\lambda)|\Psi_r(\lambda)}|\,.
\end{equation}
Although we define the simulator fidelity for any possible target state, we focus on the ground state.
As Fig.~\ref{FigureStatics}B shows, this overlap is considerably suppressed near the QPT, reaching values as low as 55\% (for $r=0.1$ in a chain of $L=400$ sites). When scaling to larger systems, $F(r,\lambda)$ will typically vanish exponentially fast,
simply due to the exponential growth of the dimension of the
Hilbert space (a kind of ``orthogonality catastrophe"). 
In a universal quantum computation, the fidelity would have to be very close to 1 for the quantum computer to work fault-tolerantly.
However, QSs have the advantage that we do not necessarily demand of the entire state to be robust.
Often, it is enough if we can distinguish the relevant phases by measuring faithfully \emph{local} observables (local in  the quantum information sense that few sites are involved, although they may be physically far apart).
Obviously, this is less demanding, yet very useful.

To quantify the robustness of local observables, we investigate
the single-site and (nearest-neighbors) two-site simulator fidelity $f_1(r,\lambda)$ and $f_2(r,\lambda)$, respectively. As the one- and two-particle density matrices will generally be mixed when the overall pure many-body state is entangled, these are defined as the Uhlmann fidelity \cite{Uhlmann1976} between the single- or two-site reduced density matrices of the ideal state and the one at disorder strength $r$, $f \equiv \mathrm{Tr}\sqrt{\sqrt{\rho_0}\rho_r\sqrt{\rho_0}}$.
It can be assumed that fidelities of the reduced system decrease more or less monotonically with the number of sites involved.
As seen in Fig.~\ref{FigureStatics}B, the reduced simulator fidelities are much more robust to disorder than the global one -- near the phase transition, $f_2(r,\lambda)$ decreases to approximately 0.998, and $f_1(r,\lambda)$ remains above 0.999.
This gives optimism that local quantities are robust enough to allow a faithful distinction between different quantum phases.

One step beyond local properties of the ground state are the \emph{correlation lengths} dictating the exponential decay of long-distance correlation functions.
We investigate the correlation length $\xi$ extracted from the correlation function
\begin{equation}
C(i,j)=\langle \Psi_r | \sigma_{z}^{(i)} \sigma_{z}^{(j)}| \Psi_r
\rangle- \langle \Psi_r | \sigma_{z}^{(i)}| \Psi_r \rangle \langle
\Psi_r | \sigma_{z}^{(j)}| \Psi_r \rangle\,,
\end{equation}
where away from criticality $C(i,j)\propto \exp(-\left|i-j\right|/\xi)$.
Without disorder and for infinite systems, $\xi$ diverges at the critical point, because criticality is the emergence of collective phenomena involving infinite degrees of freedom at all length scales. 
In practice, we can only deal with finite systems so that we cannot observe real criticality but only smoothed out signatures of it, a phenomenon which one normally calls ``pseudo-criticality.'' For example, the correlation length $\xi$ is bounded by the system size. Still, its peak gives a reliable signature for the location of the critical point. 
As Fig.~\ref{FigureStatics}A shows, however, disorder suppresses correlations and broadens the peak of $\xi$, thus making an extraction of the critical point much less reliable.

Another criterion to locate the QPT is provided by the \emph{energy gap} $\Delta$ between ground state and first excited state. At criticality, the low-energy spectrum of the Hamiltonian is gapless in the thermodynamic limit. In finite systems, it presents non-vanishing gaps that decrease in a systematic way with increasing system size. Due to this characteristic scaling of physical observables as a function of system size in pseudo-critical systems, criticality can be detected by studying a sequence of finite but increasingly large systems, a technique called \emph{finite-size scaling} \cite{Barber1983}. 
We describe this technique for the energy gap in the Appendix (see Fig.~\ref{FigureGapCrossings}), and show in Fig.~\ref{FigureStatics}A (black line) the location of the critical point extrapolated in this way. As can be seen, 
if one does not correct for disorder effects, one would locate the critical point at values of $\lambda$ that are too large.

Perhaps of more fundamental interest than the exact location of a critical point is its \emph{universality class}.  
All models within a given universality class give rise to the same collective behavior at large distances (typically large with respect to the lattice spacing), irrespective of their microscopic details~\cite{bookCardy}. Therefore, all relevant thermodynamic quantities for all models within a class are characterized by the same small set of \emph{critical exponents} which describe the power-law decay of the correlation functions of local observables in the large-distance regime, a property that allows to differentiate among different emerging collective behaviors.
To investigate how robust the universal behavior is, we compute the critical exponent for the correlation length, $\nu$,
from a collapse of the correlations (as explained in the Appendix, see Eq.~\eqref{eq:collapse} and Fig.~\ref{FigureCollapse}).
As shown in Fig.~\ref{FigureStatics}C, already for a few percent of disorder, $\nu$ increases strongly from its ideal value 1.
Therefore, if one simply neglects the influence of disorder, the extraction of critical exponents yields wrong results.

If the QPT is described by a CFT (a specific subclass of one-dimensional critical systems), it is characterized by a \emph{central charge} $c$.
The central charge appears ubiquitously \cite{CardyLecture}. It, e.g., governs the temperature dependence of the
free energy (Stefan-Boltzmann law) and the Casimir effect in finite geometries, but also the scaling of the entanglement entropy of sub-regions of the ground state of the corresponding quantum models. Models with different central charge have different emerging collective behavior. For example, models whose collective behavior is that of a free Majorana fermion (as in the disorder-free TIM) have central charge $=1/2$, while models whose collective behavior is that of a free boson have central charge $=1$. 
Strictly speaking, the TIM has an underlying CFT only in the disorder-free case, but there have been efforts to extract an effective central charge also for the disordered model~\cite{Refael2004}, for example, from the von Neumann entropy $S$ of the reduced density matrix of a part of the chain of size $l$.
At criticality, this entropy scales as $c/6 \log(L/\pi \sin(l\pi/L)) + A$
\cite{Holzhey1994,Vidal2003,Calabrese2004}.
Figure \ref{FigureStatics}C shows that disorder decreases the effective central charge.
Hence, ignoring the effects of disorder would give completely erroneous results, since even a small deviation of the central charge indicates a completely different universality class.
Note also that the decrease of $c$ with disorder indicates the destruction of correlations by disorder.

Fortunately, for all the extracted quantities (except the global simulator fidelity), the levels of disorder for appreciable changes to occur are at least a few percent.
If the AQS can be operated below such a value, its results seem to be robust, at least in this simple model system.
State-of-the-art experiments are good enough to fulfill this requirement.
In fact, in many experimental situations one hopes to reach  levels of disorder or
noise that are below a few percents, assuring the robustness of
the AQS. Frequently, however, changing parameters from the regime
where validation via classical simulation is possible to the
regime of {\it terra incognita} might lead to uncontrolled
disorder or noise. That is why checking sensitivity to disorder
and noise in those regimes where it can be checked is of great importance.

\vskip 0.5cm

\section{Results -- Dynamics}

Efficient classical algorithms for computing static properties of
quantum systems are  more developed than for computing dynamics
(the difficulty arises mainly because entropy and correlations
grow rapidly with simulated time). Therefore, one can assume that
in the absence of disorder, a quantum simulation of dynamics can
much more easily outperform classical computers. Indeed, in a
recent experiment based on ultracold bosonic atoms, the controlled
dynamics ran for longer times than present classical algorithms
based on matrix product states could efficiently track
~\cite{Trotzky2012}. We thus turn to the issue of how disorder
affects the reliability of quantum simulations of dynamics. As
with statics, we investigate the behavior of the simulator
fidelity, but now also as a function of time, initial state, and
external driving.

Typically, we expect that the simulator fidelity will decay with time, and eventually reach an asymptotic
finite value. The effect of disorder in both the decay rate and the asymptotic saturation value can, in general,
be understood from established techniques such as Fermi Golden's rule, and random matrices~\cite{Gorin2006}.
On the other hand, the effect of the initial state and the external driving is known to be nontrivial and
of particular interest for our purposes. For example, it is known that numerical techniques such as the time-dependent
density matrix renormalization group (tDMRG) can simulate efficiently the dynamics after a sudden quench of the
field $h$, as long as the quench is restricted to a few sites on the chain. However, if the quench is global,
it has been shown that the computational resources needed to keep a fixed amount of error grow
exponentially with time~\cite{Prosen2007,Perales2008}.
Generically, solving for the dynamics of a quantum many-body system is a hard problem 
for classical algorithms. Our model is special because it can be solved exactly for all cases, although it remains hard
for the tDMRG algorithm. We use this to our advantage to study how this class of classical algorithms
behaves when solving for quantum dynamics.

\begin{figure}
\centering
\includegraphics*[width=8.6cm]{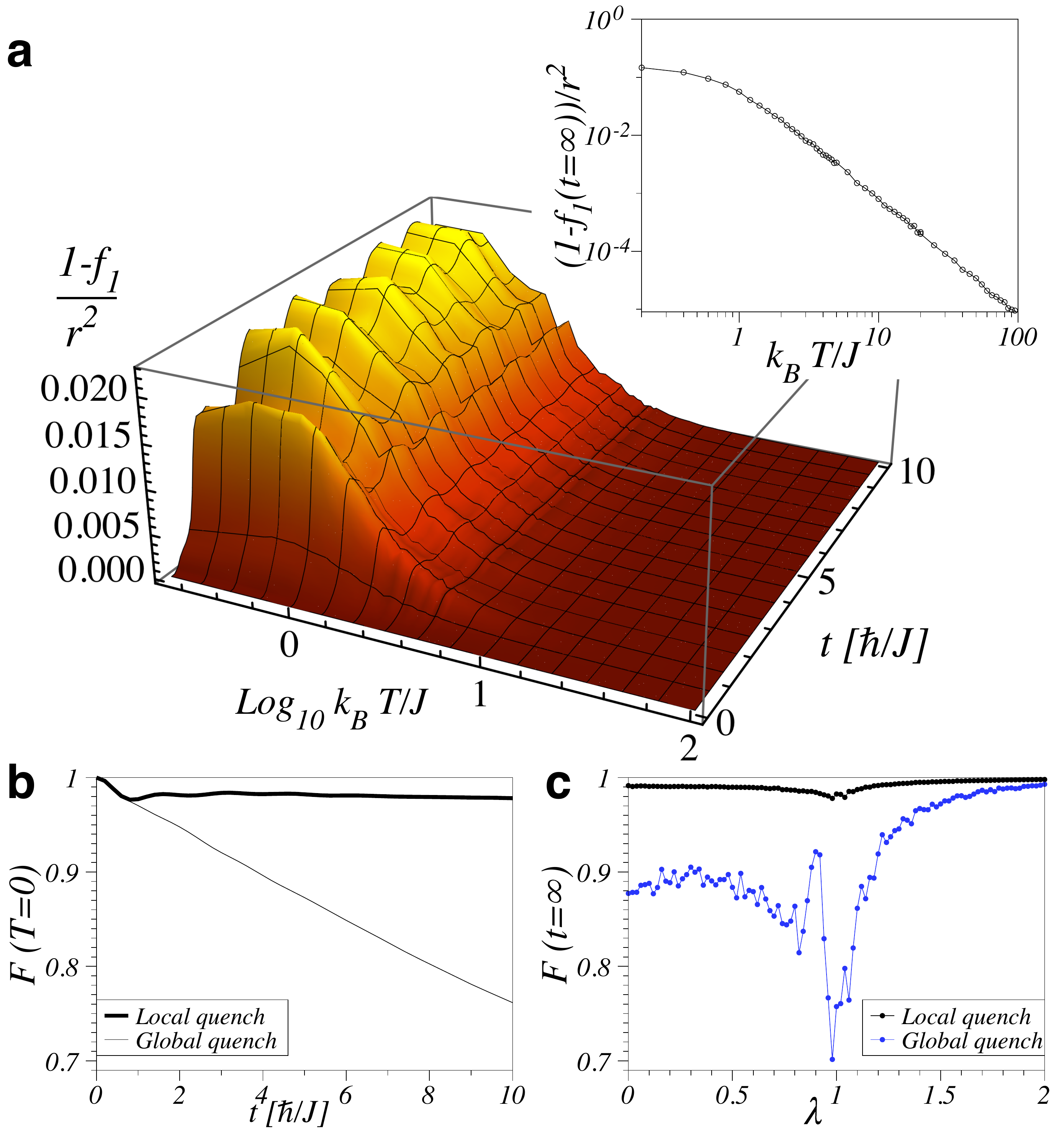}
\vskip-0.4cm \caption{
\textbf{A}: Evolution of the average reduced
simulator fidelity as a function of the temperature of the initial
state. The system is an Ising spin chain of length $50$, the initial state is a thermal state at criticality ($\lambda=1$),
and at time zero the field is suddenly quenched to $\lambda=2$. In
the vertical axis we show the infidelity (one minus fidelity)
normalized by the disorder strength $r$ squared.
For larger temperatures (where there are less correlations) the state is more robust.
In the inset, we
show the average asymptotic infidelity as a function of
temperature. For large temperatures it decays as
$1/T^2$.
\textbf{B:} Evolution of the full simulator fidelity for an initial state equal
to the ground state (zero temperature) at $\lambda=0.75$ after a sudden quench to
$\lambda=1.25$.
For a local quench in a single site, fidelity saturates rapidly at large values, but decreases strongly for a global quench.
\textbf{C:} Asymptotic value of the total simulator fidelity as a function of the initial value of the field
$\lambda$, with a fixed quench strength of $\delta\lambda=0.25$.
The system is much less robust for global quenches and near criticality ($\lambda=1$).
}
 \label{FigureDynamics}
\end{figure}

We studied the behavior of the full simulator fidelity under the following driving. As initial state we prepare the ground state of the Hamiltonian for a given value of the external field.
At time zero, the field is quenched instantaneously to a larger strength, and the system is allowed to evolve. In panels~B and~C of Fig.~\ref{FigureDynamics}, 
we compare the short- and long-time behavior of fidelity for the case of a global and a local (single-site) quench. 
The AQS keeps a high fidelity in the case of a local quench, while it  performs poorly for the global quench, with fidelities reaching lows of $0.8$ 
even for small systems of $50$ spins. We also observe that the AQS performs worse when the  quench crosses the critical point, 
as shown in Fig.~\ref{FigureDynamics}C, where we fix the strength of the quench and vary the initial field value.

The initial state can also have an effect on the efficiency of classical algorithms. Using the same setup with a global quench,
but starting from a thermal initial state, tDMRG becomes efficient for high temperatures~\cite{Prosen2007}
where the state and its correlations are almost
classical. However, it becomes exponentially inefficient with time for low-temperature initial states.
For initial thermal states, we can still compute the dynamics exactly, although computationally it becomes too expensive to calculate the
full many-body fidelity between the evolved states.
In this case, therefore, we focus on the reduced simulator fidelity. For the regimes of disorder that
we studied, we observe that the time dependent fidelity decays with a rate roughly proportional to the strength
of the disorder squared
(typical of a Fermi golden rule \cite{Gorin2006}). 
For this reason, we show in Fig.~\ref{FigureDynamics}A a rescaled form of the fidelity, $(1-f_1)/r^2$,
that exemplifies the typical behavior for all disorder strengths, as a function of time and temperature of the
initial state.

As with the classical algorithms~\cite{Prosen2007}, the AQS remains faithful when the state is almost classical (high
temperatures). The simulator fidelity decreases rapidly for low temperatures, although it saturates at a fairly high value.  In terms of distinguishability, the values we find imply that a fair observer would have only a $4\%$ chance of
distinguishing the 1-spin reduced state of the AQS from the ideal state.
In the inset of the top panel we show the average asymptotic fidelity as a function of temperature of the initial state.
Again, for low temperatures fidelity worsens, but saturates to a few percent. For high temperatures, it is simple to perform an expansion of the fidelity which shows that $f_1 \simeq 1-T^{-2}$.

\section{Discussion and Outlook} 
A key issue for future investigation is the relationship between the robustness of an
analog quantum simulator and its computational power.  For the
models we have considered here, the physically relevant
correlation functions are robust for a reasonable degree of
disorder.  This suggests that such an AQS could perform well in a laboratory demonstration. But, the TIM that we
considered here is simulatable on a classical computer. Is this
connection between robustness and classical simulatability
coincidental, or does it reflect a deeper relationship?

Disorder reduces the correlation length of the spin chain. Because
less-correlated quantum states can be described with fewer
parameters, there is reason to suspect that certain aspects of
weakly disordered quantum many-body systems could actually be {\em
easier} to simulate on classical computers than their clean
idealized versions. This happens, for example, in the realm of
digital quantum computation, where a quantum circuit becomes
classically simulatable for noise levels above a certain level
when quantum gates lose their entangling power~\cite{Aharonov1996,
Harrow2003, Virmani2005}.  In the context of many-body physics, the
success of DMRG in, e.g., 1D spin chains, is rigorously related to
the existence of efficient matrix-product-state
representations~\cite{Vidal2003a}. These take advantage of the
small amount of quantum correlations in such systems, thus
compressing the $O(\exp(n))$ parameters needed to describe a
general $n$-particle state to $O(n)$ finite-dimensional matrices.
In higher-dimensional lattices,  states which obey the so-called
``Area law''~\cite{Eisert2010}, where quantum correlations are smaller
than in generic states,  may still be amenable to a classical
simulation using state-of-the-art techniques such as tensor
networks~\cite{Verstraete2004,Tagliacozzo2009,Tagliacozzo2011,Evenbly2011a,Schollwoeck2011}, Density Functional
Theories~\cite{Martin2004}, or Quantum Monte Carlo~\cite{Sandvik2010a}.

We thus arrive at the fundamental question: {\em Do the finite
imperfections of an analog quantum simulator reduce the
correlations, and thus the number of parameters needed to describe
the system, so as to render the device simulatable by classical
means}? We know that for noise above certain levels a digital
quantum circuit is classically simulatable and for levels below a
certain threshold it can be rendered fault tolerant.  Is there an
{\em intermediate regime} for which noise is too great to allow
fault-tolerant universal quantum computation, but small enough
that an AQS accesses physics beyond classical
simulation?  The existence of an intermediate regime would imply
that there exists a whole class of problems outside $P$ that we
can access in the near future, even without a fully functioning
quantum computer.

The results we present here, in particular those for dynamics, are
an initial attempt -- albeit in a trivial model -- at
understanding the above problem.  We can see how an analog quantum
simulator works well when a classical solution is efficient, and
worsens (but only in a limited way) when the problem becomes
classically hard to simulate. Even though the underlying model is
actually solvable,  this may be positive evidence for the
existence of an intermediate regime of noise, and the efficiency
of AQSs in more complex situations.

Our main discussion focused on AQSs, but similar issues pertain to DQSs.
Since to date there exists no known way to fault-tolerantly
error-correct AQSs, there is a natural tendency to explore the advantages of DQSs, where error correction is possible. 
The above discussion shows, however, that a digital implementation of a
quantum simulation does not, in itself, guarantee an efficient and
more powerful simulation than one that is carried out classically.
As in any quantum algorithm, initialization, evolution of the
state, and measurement must be performed efficiently, i.e., with a
polynomial use of physical resources (space and time). 
Digital quantum simulation is no exception.  Indeed, as discussed above, a fault-tolerant implementation of the standard approach based on the Trotter expansion~\cite{Lloyd1996} comes at the cost of
an overhead in the number of gates and time required that grows exponentially with the degree of precision required ~\cite{Brown2006,Clark2009}.
If we can guarantee the reliability of analog quantum simulators while avoiding such exponential costs, many open problems from all areas of physics could suddenly come into the reach of being solved.

Finally, we can turn the problem of quantum simulation on its head and ask, what does Nature do?  For any real material, like a high-$T_c$ cuprate, has imperfections.   Does Nature access highly correlated states that cannot be efficiently simulated on a classical computer?   Certainly, in some cases we believe it does, as for example in high-$T_c$ superconductors~\cite{Lee2008} or in certain ground states of frustrated quantum antiferromagnets which are believed to carry topological order~\cite{Misguich2004}. If noise is low enough, does Nature protect quantum correlations to a degree that classical methods cannot efficiently represent the physically interesting quantities? And, can we exploit this capability with a quantum simulator? If Nature does it, we should take advantage of it!

\vskip 0.5cm

\noindent{\bf Acknowledgements ---}
We gratefully acknowledge support by the Caixa Manresa, Spanish MICINN (FIS2008-00784 and Consolider QOIT), AAII-Hubbard, EU Projects AQUTE and NAMEQUAM, ERC Grant QUAGATUA, and Marie Curie project FP7-PEOPLE-2010-IIF ``ENGAGES'' 273524.  IHD was supported by NSF grants 0969997 and 0903953.  We also acknowledge fruitful discussions with Carl Caves, Dave Bacon, Robin Blume-Kohout, Rolando Somma, and Roman Schmied.


\vskip 1cm

\noindent{\large\bf{Appendix}}

\renewcommand{\theequation}{A\arabic{equation}}
\setcounter{equation}{0}
\renewcommand{\thefigure}{A\arabic{figure}}
\setcounter{figure}{0}

\vskip 0.5cm

\noindent{\bf Quadratic fermionic systems ---}
The transverse field Ising model, Eq.~(1) of the main text, even with disorder, can be solved by casting it into the form of non-interacting fermionic particles
using the Jordan--Wigner transformation,
\begin{subequations}
\eqa{
\sigma_j^+&=&c_j^\dagger \prod_{m=1}^{j-1} \ue^{-i \pi c_m^\dagger c_m} \,, \\
\sigma_j^-&=&\prod_{m=1}^{j-1} \ue^{i \pi c_m^\dagger c_m} c_j \,, \\
\sigma_j^z&=& 2 c_j^\dagger c_j - 1 \,.
}
\end{subequations}
The $c_j$, $c_j^\dagger$ obey on commutation relations.
This transformation leads to
\eq{
\label{eq:HJW}
\hat H=\sum_{i,j}  \left[ c_i^\dagger A_{ij} c_{j} + \left(c_i^\dagger B_{ij} c_{j}^\dagger +h.c.\right)\right] - \frac{1}{2} \sum_{j} A_{jj} \,,
}
where
\begin{subequations}
\eqa{
A_{ij}&=&-J_{ij}\left(\delta_{j,i+1} + \delta_{j,i-1} \right) - 2  h_i \delta_{j,i}\,, \\
B_{ij}&=&-J_{ij}\gamma\left(\delta_{j,i+1} - \delta_{j,i-1} \right)\,.
}
\end{subequations}
Hamiltonian \eqref{eq:HJW} can be diagonalized to
\eq{
\label{eq:HJWdiag}
\hat H=\sum_{k=1}^N \Lambda_k \eta_k^\dagger \eta_k + E_0\,,
}
where $\Lambda=\Phi\left(A-B\right)\Psi^\intercal$ is diagonal. $\Lambda$, $\Phi$, and $\Psi$ can be obtained from the singular-value decomposition of $Z\equiv A-B$.
The normal modes are $\eta_k=\sum_{j=1}^N\left(g_{k,j} c_j + h_{k,j} c_j^\dagger\right)$, where $g=\left(\Phi+\Psi\right)/2$, and $h=\left(\Phi-\Psi\right)/2$.
From this, we can compute the relevant ground-state properties.

\vskip 0.5cm

\noindent{\bf Ground-state fidelity and correlations ---}
From the normal modes obtained in the diagonalization of the previous section, we can compute the observables we are interested in:
the simulator fidelity $F$ (the overlap to the disorder-free ground state), reduced simulator fidelities, the energy gap, and the $ZZ$-correlations.

In general, the overlap between the ground states of two realizations $Z$ and $\widetilde{Z}$ is~\cite{Cozzini2007}
\eq{
\label{eq:FZZtilde}
F\left(Z,\widetilde{Z}\right)=\sqrt{\det \frac{1+T^{-1}\widetilde{T}}{2}}\,,
}
with $T=\left(\Phi^{-1}\Lambda\Phi\right)^{-1}Z$.
We define the simulator fidelity $F$ as the overlap at fixed $\lambda$ between the ideal, disorder-free state and the state at disorder strength $r$,
\eq{
\label{eq:F0}
F(r,\lambda)\equiv F\left(Z(\lambda)_r,Z(\lambda)_{0}\right)\,.
}
This is a global quantity, but one can expect that local observables are less affected by disorder.
A measure for the change of local quantitites is the \emph{single-site simulator fidelity}
\eq{
\label{eq:f0single}
f_1(r,\lambda)=\sum_{i=1}^L\mathrm{tr}\sqrt{\sqrt{\rho_0^{(i)}(\lambda)}\rho_r^{(i)}(\lambda)\sqrt{\rho_0^{(i)}(\lambda)}}\,,
}
where $\rho_r^{(i)}=\mathrm{tr}_{j\neq i} \rho$ is the reduced density matrix of site $i$ under disorder $r$, and $\rho_0^{(i)}$ is the equivalent in the disorder-free case.
The single-site reduced density matrix is completely determined by the expectation values of $\sigma_i^{\mu}$, $\mu=x,y,z$, since one can expand $\rho^{(i)}=\frac{1}{2}\sum_{\mu}\braket{\sigma^\mu_i}\sigma^\mu$. Here, the sum runs over $\sigma^{\mu}$, $\mu=x,y,z$, and $\sigma^{(0)}=\mathbb{1}$.
We also analyse the \emph{two-site simulator fidelity}
\eq{
\label{eq:f0two}
f_2=\sum_{i=1}^L\mathrm{tr}\sqrt{\sqrt{\rho_0^{(i,i+1)}(\lambda)}\rho_r^{(i,i+1)}(\lambda)\sqrt{\rho_0^{(i,i+1)}(\lambda)}}\,,
}
for nearest neighbors.
Here, $\rho_r^{(i,i+1)}=\mathrm{tr}_{j\neq i,i+1} \rho$ is the reduced density matrix of sites $(i,i+1)$ under disorder $r$, and $\rho_0^{(i,i+1)}$ is the equivalent in the disorder-free case.
We compute all considered static quantities as the mean over a large number of disorder realizations;
for the fidelities $F$,$f_1$, and $f_2$ displayed in Fig.~1B of the main text, we used 5000 realizations at chain length $L=400$.

The correlations, finally, can be computed using the fact that the ground state $\ket{\Psi}$ of Eq.~\eqref{eq:HJWdiag} is the vacuum of the normal modes (\textit{i.e.}, $\eta_k\ket{\Psi}=0$, $\forall k$). For example, for the $ZZ$-correlations this yields
\eqa{
C(i,j)&\equiv&\langle \Psi_r | \sigma_{z}^{(i)} \sigma_{z}^{(j)}| \Psi_r
\rangle- \langle \Psi_r | \sigma_{z}^{(i)}| \Psi_r \rangle \langle
\Psi_r | \sigma_{z}^{(j)}| \Psi_r \rangle\nonumber \\
    &=&4\langle \Psi_r |c_i^\dagger c_i c_j^\dagger c_j| \Psi_r \rangle
        -4\langle \Psi_r |c_i^\dagger c_i| \Psi_r \rangle \langle \Psi_r |c_j^\dagger c_j| \Psi_r \rangle \nonumber \\
    &=& 4 \left(h^\intercal h\right)_{ij} \left(g^\intercal g\right)_{ij}
        - 4 \left(h^\intercal g\right)_{ij} \left(g^\intercal h\right)_{ij} \,.
}
Away from criticality, the correlations decay as $C(i,j)\propto \exp(-\left|i-j\right|/\xi)$ with \emph{correlation length} $\xi$.
In Fig.~1A of the main text, we display $\xi$ extracted from fits to part of the wings of $C(i,j)$ (for $L=400$ and 10000 disorder realizations).

\begin{figure}
\begin{center}
\includegraphics*[width=4.3cm]{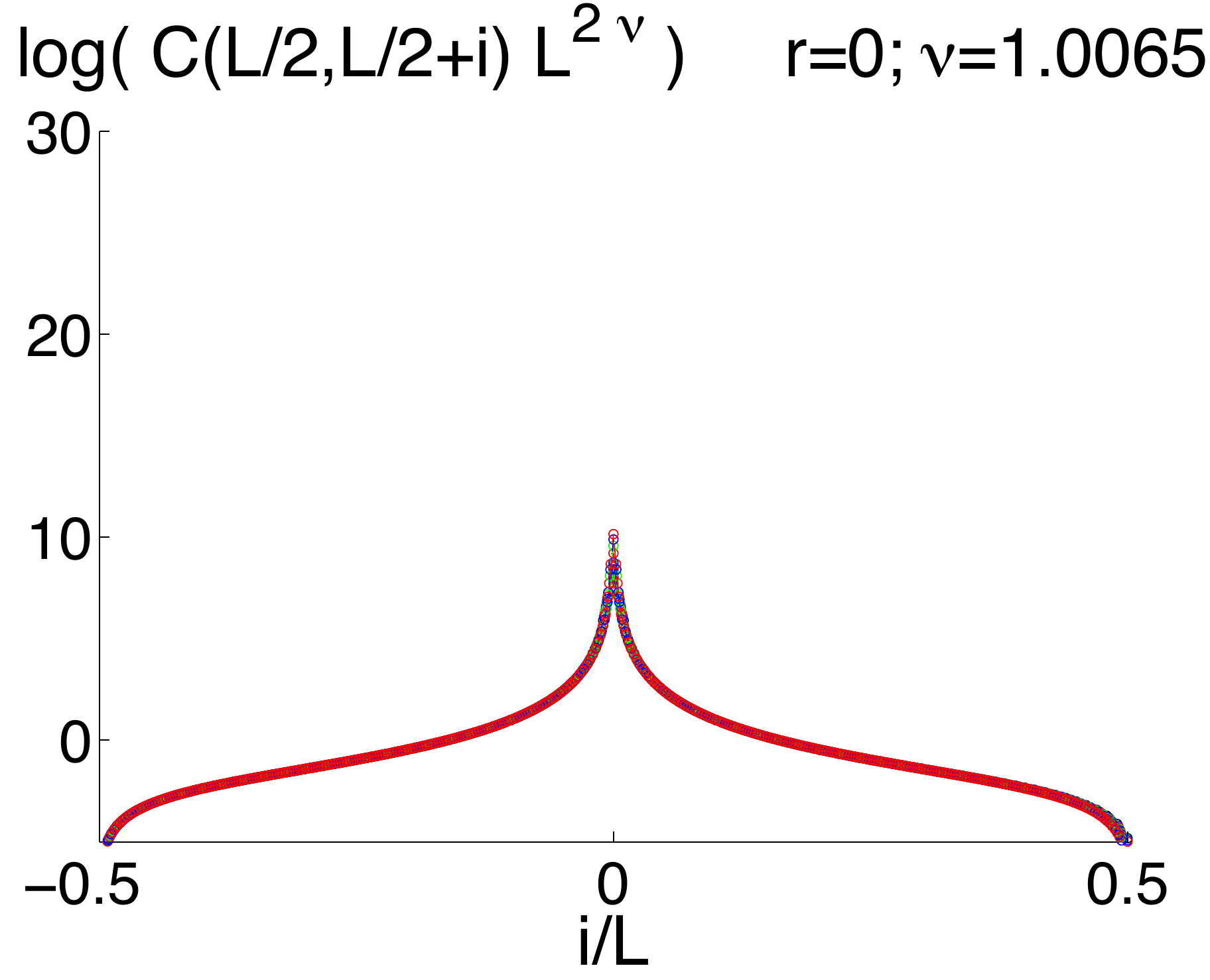}\includegraphics*[width=4.3cm]{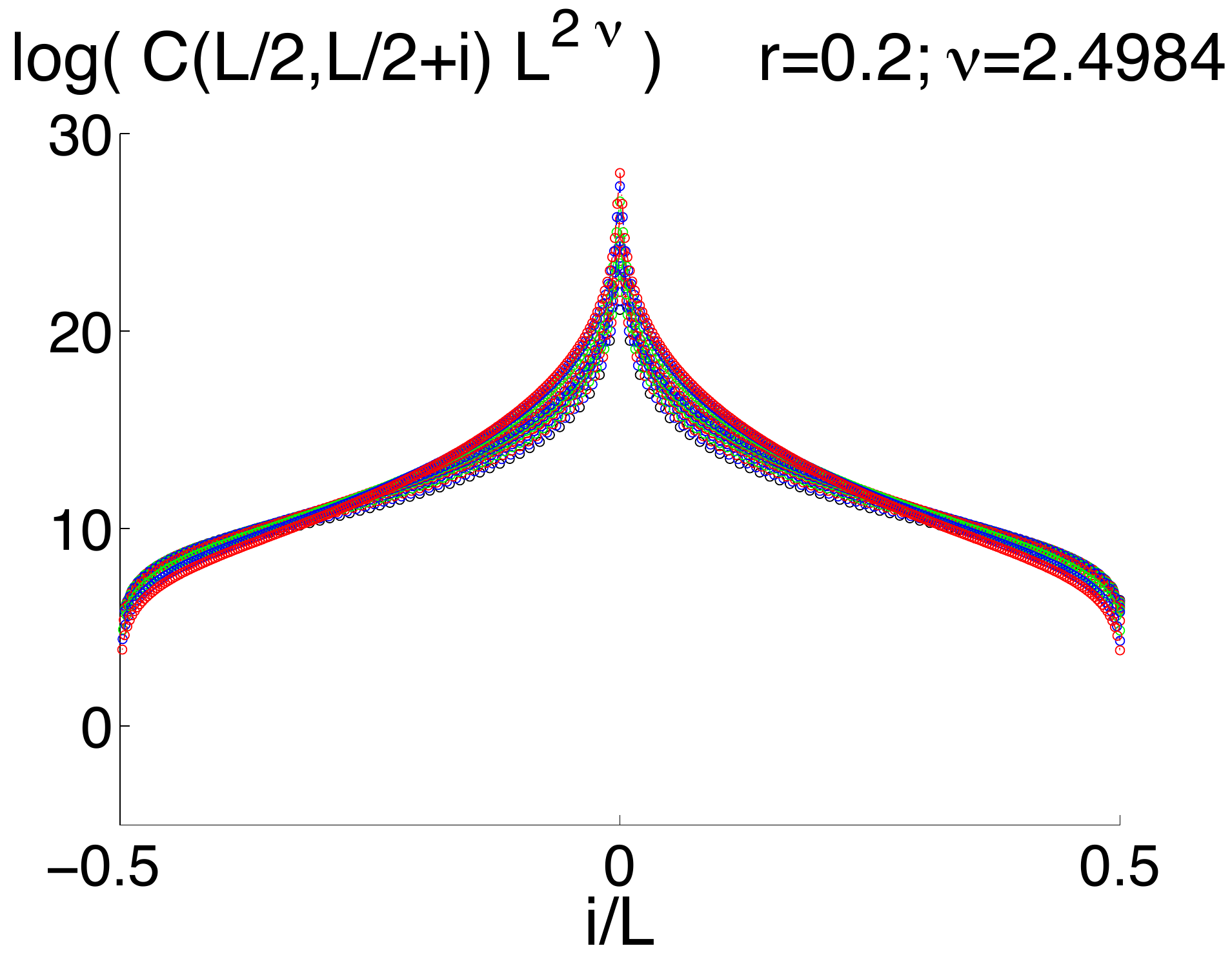}\\
\vskip0.4cm
\includegraphics*[width=4.3cm]{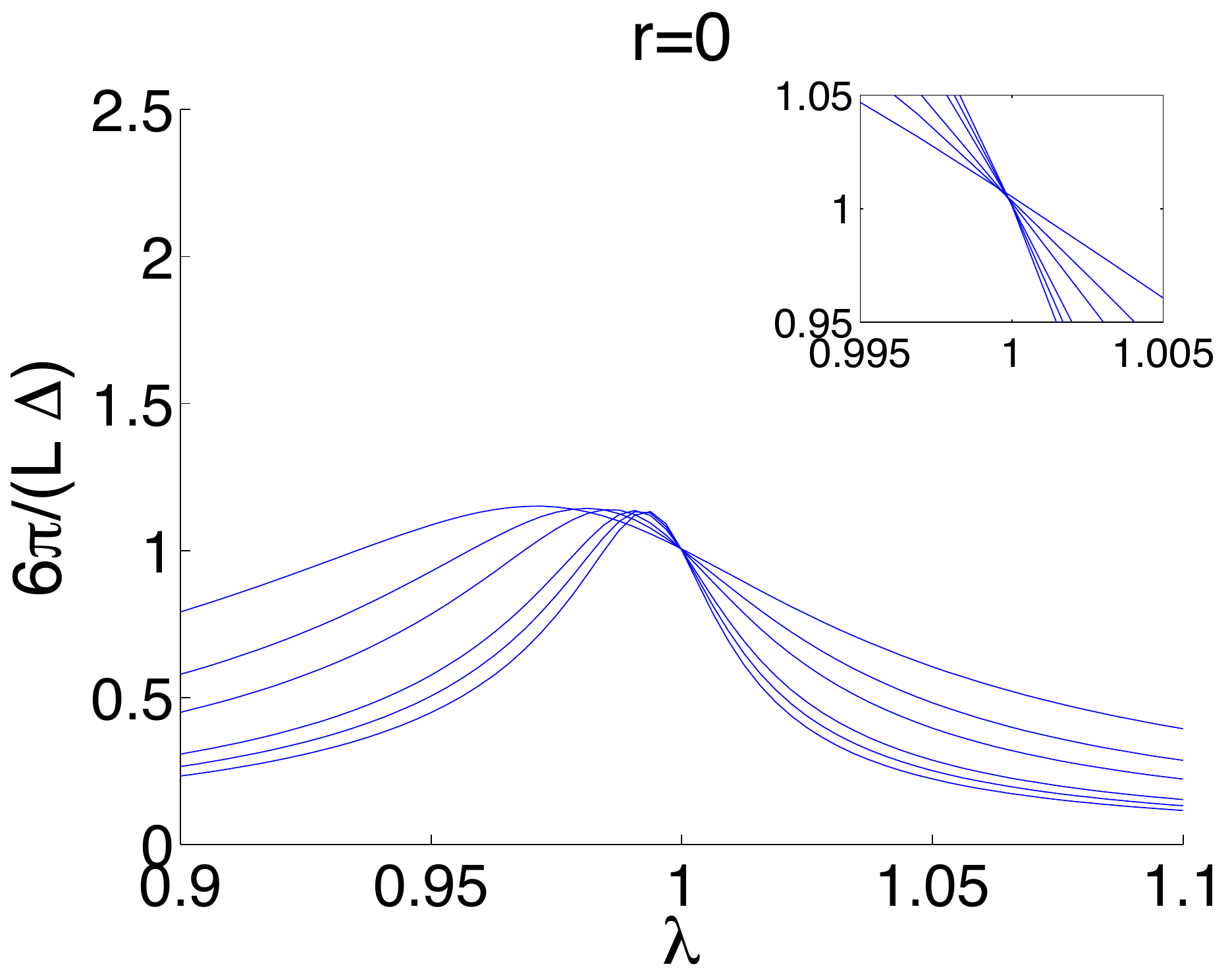}\includegraphics*[width=4.3cm]{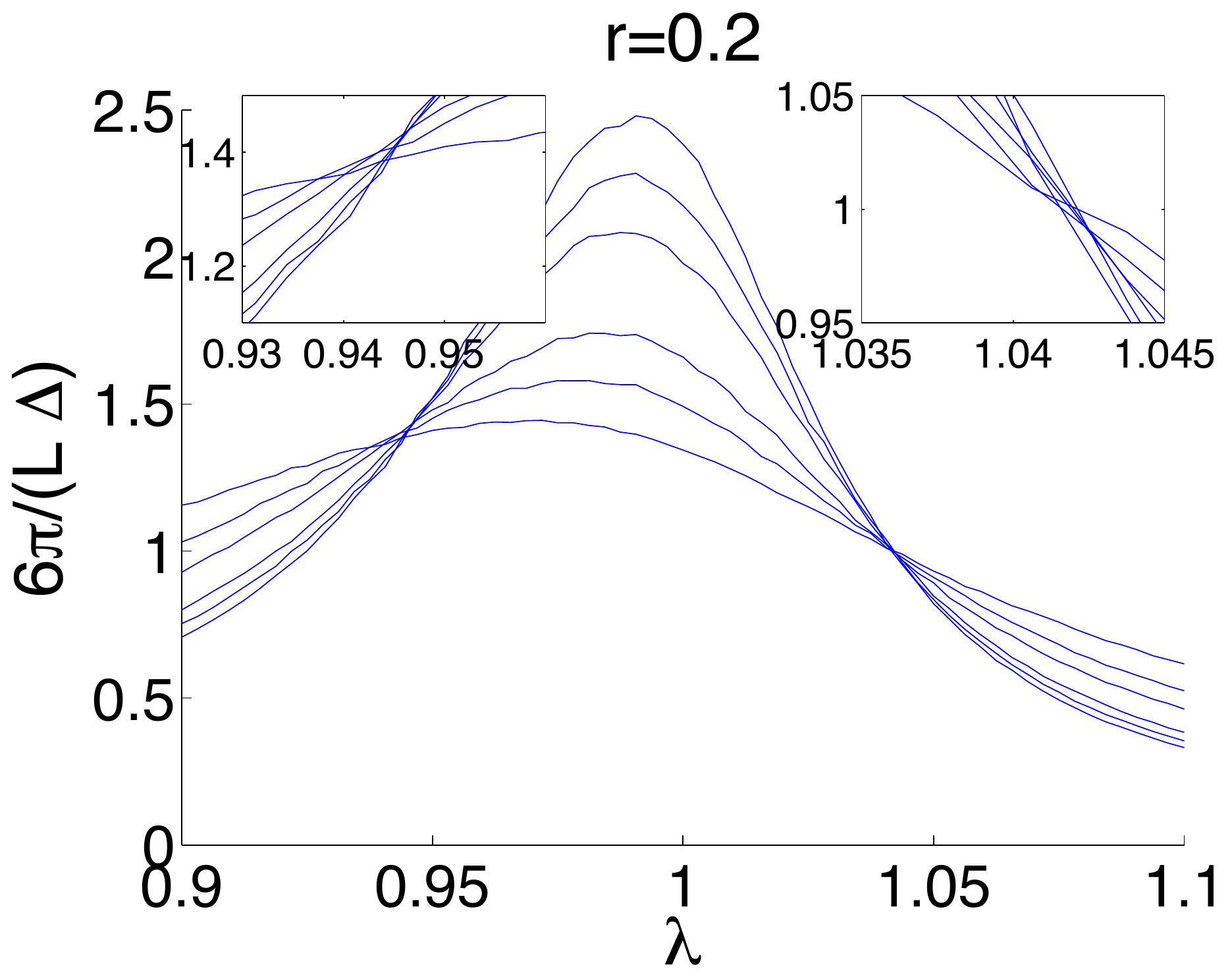}
\end{center}
 \caption{
 \textbf{Upper row:}
 For low disorder ($r=0$, left), the curves $C(i,j) L^{2 \nu}$ for different $L$ plotted as functions of $\left|i-j\right|/L$ collapse perfectly for the physically correct value of the critical exponent $\nu=1$ (dots of different color correspond to different $L$).
 For increasing disorder ($r=0.2$, right), the collapse worsens, and the best collapse is obtained for some $\nu>1$.
 \textbf{Lower row:}
 For low disorder ($r=0$, left), the curves $1/(L \Delta(L))$ cross perfectly at the location of the critical point, $\lambda=1$.
 With increasing disorder ($r=0.2$, right), the crossing point moves to larger values of $\lambda$ and becomes less well defined.
 Also, at large disorder there appears a second crossing point below $\lambda=1$. The insets show zooms on the crossing points.
 The findings of both rows of figures mean that, ignoring the effects of disorder in an imperfect AQS yields, compared to the ideal model, too large values for the critical exponent as well as the location of the critical point.
 }
\label{FigureGapCrossings} \label{FigureCollapse}
\end{figure}

Without disorder,
\begin{equation}
    \label{eq:collapse}
    C(i,j) L^{2 \nu} \propto f(\left|i-j\right|/L)
\end{equation} for some universal function $f$~\cite{bookCardy}. Hence, one can extract the critical exponent for the correlation length $\nu$ from a data collapse of the correlations.
In Fig.~1C of the main text, we show the erroneous values for $\nu$, extracted from Eq.~\eqref{eq:collapse} if one naively neglects that this relationship is no longer true in the presence of disorder.
Figure~\ref{FigureCollapse} shows the best collapse achieved with Eq.~\eqref{eq:collapse} for disorder levels $r=0$ and $0.2$.
The value of $\nu$ for the best collapse increases with disorder. Hence, using Eq.~\eqref{eq:collapse} on a disordered AQS yields a too large critical exponent, compared to the ideal model.
Moreover, the qualitity of the collapse worsens with increasing disorder, demonstrating that a naive application of Eq.~\eqref{eq:collapse} is unjustified if disorder is large.
For this analysis, we used $L=100$ to $190$ in steps of $10$ with $10^6$ disorder realizations, $L=200$, $250$, and $300$ with $5\times10^5$ realizations, and $L=350$ and $400$ with $10^5$ realizations.

The correlations are intrinsically connected to the energy gap $\Delta(L)$, since a gapped system necessarily has exponentially decaying correlations~\cite{Cramer2006}.
Via a finite-size scaling, the gaps at finite systems also allows to extract the location of the QPT in the infinite system, as seen in Fig.~\ref{FigureGapCrossings}. There, we plot curves $1/(L^{\zeta} \Delta(L))$ for closeby chain lengths $L$, where $\zeta$ is the dynamical critical exponent, which for the disorder-free case equals 1. These cross at a series of pseudo-critical points which with increasing $L$ tends rapidly to the critical point of the thermodynamic limit~\cite{Nightingale1975}. 
Assuming that it does not change much for small disorder, we identify as an approximation the critical point with the mean of the crossing points curves $1/(L \Delta(L))$ for $L=100,150,200,300,350,400$ with 10000 realizations of disorder each.
As displayed in Fig.~\ref{FigureGapCrossings}, the crossing point moves to larger values of $\lambda$ with increasing disorder (see also Fig.~1A of the main text).
This means that applying this analysis to a real-world AQS without correcting for disorder can yield erroneous results (compared to the ideal model) for the location of the QPT.
Moreover, the crossing point becomes less well defined with increasing disorder showing that this analysis should be corrected for the presence of disorder.
Finally, at large disorder, a second crossing point appears at lower $\lambda$. The two crossing points open up to a V-like structure with increasing disorder.
This could be interpreted as an indication of the Griffiths phase (the crossing points are qualitatively consistent with the extent of the Griffiths phase found in~\cite{Jacobson2009}). For a more quantitative analysis, however, one would need to account for a change of $\zeta$ with increasing disorder.

If a given model can be described by a conformal field theory, its universal critical behavior (including all critical exponents) is completely defined by a single number, the central charge $c$.
To extract it, we compute the von Neumann entropy $S$ of the reduced density matrix of a part of the chain of size $l$ for $L=300$ with 10000 disorder realizations.
For systems with open boundary conditions, a fit to $c/6 \log(L/\pi \sin(l\pi/L)) + A$ (excluding small values of $l$) yields the effective central charge $c$~\cite{Holzhey1994,Vidal2003,Calabrese2004}. The disorder-free value is $c=0.5$. Increasing disorder suppresses this, indicating the decrease of entanglement in the system (see Fig.~1C of the main text).
Again, applying the analysis that is correct in the disorder-free case (where the system is indeed described by a conformal field theory) without adjustments to the disordered system, yields results which deviate from the desired ideal case.

\vskip 0.5cm

\noindent{\bf Time dependent fidelities ---}
For time evolution, we distinguish between the zero and finite temperature fidelities, although the underlying
technique is the same. We start by rewriting the fermionic Hamiltonian above as
\be
\hat H = \frac{1}{2} \vec{\Psi}^\dagger \cdot {\mathrm H} \cdot \vec{\Psi},
\ee
where $\vec{\Psi}^\dagger = ( c^\dagger_1, ...,c^\dagger_N, c_1, ...,c_N)$ is a $2N$ length vector composed of
all creation and annihilation operators present in $\hat H$, and ${\mathrm H} = A \otimes \sigma_z + i B \otimes \sigma_y$
is a  $2N \times 2N$ matrix with complex coefficients.

For computing fidelities, we use  the convenient Levitov's formula~\cite{Levitov1996,Klich2003}, which relates traces of operators in the
Hilbert space of the fermions to determinants of much smaller matrices (like ${\mathrm H}$). For example, let
$\hat P=\vec{\Psi}^\dagger \cdot {\mathrm P} \cdot \vec{\Psi}$ and $\hat Q=\vec{\Psi}^\dagger \cdot {\mathrm Q} \cdot \vec{\Psi}$
be two operators in the space of fermions,
with ${\mathrm P}$ and ${\mathrm Q}$ complex valued $2N \times 2N$ matrices. Then,
\be
{\mathrm{Tr}} \left( e^{\hat P} e^{\hat Q} \right) = {\mathrm{det}} \left( 1 + e^{\mathrm P} e^{\mathrm Q} \right).
\ee
Similar formulas hold for more or less operators.

In the zero-temperature case,
when the initial state remains pure after evolution, the fidelity takes the form of an overlap
\be
F=|\braket{\psi_{0}(t)|\psi_{r}(t)}|,
\ee
where $\ket{\psi_0(t)}=e^{-i {\hat H}_0 t} \ket{\psi_0}$ is the initial state evolved with the target Hamiltonian of the simulation,
${\hat H}_0$, and
$\ket{\psi_r(t)}=e^{-i {\hat H}_r t} \ket{\psi_0}$ is the same state evolved with an imperfect Hamiltonian
${\hat H}_r={\hat H}_0+r {\hat V}$. Rewriting the fidelity,
\ba
F={\mathrm {Tr}} \ \rho_0 e^{i {\hat H}_r t} e^{-i {\hat H}_0 t},
\ea
with $\rho_0=\ket{\psi_0}\bra{\psi_0}$, we can use Levitov's formula and obtain
\be
F={\mathrm{det}} \left(1-G_0+G_0 e^{i {\mathrm H}_r t} e^{-i {\mathrm H}_0 t} \right),
\ee
with $G_0=\bra{\psi_0} G \ket{\psi_0}$, and G the correlation matrix of the original fermionic
operators, $G_{i,j} =  \Psi_i^\dagger \Psi_j$.

If the initial state is not pure, but a thermal state, the state remains mixed even if the evolution is unitary. In this case,
we cannot compute the fidelity for the full many-body state, but only the fidelity of the reduced density matrix for
a few spins. For this we must evaluate the correlation functions of the Pauli operators at different sites of the chain.
For the case of a single spin, the symmetry of the system ensures that at all times the reduced density matrix can be
written as $\rho = (1 + <\sigma^i_x> \sigma^i_x)/2$. Since $<\sigma^i_x> = c^\dagger_i c_i$, we only need to compute the
evolution of the diagonal terms in the $G$ correlation matrix.

\end{document}